\input ./style/arxiv-general.cfg
\documentclass[aos,MSNbibl,nameyear,dvips]{arximspdf}
\makeatletter
   \@ifpackageloaded{graphicx}{}{\usepackage{graphicx}}
\makeatother

\doi{10.1214/14-AOS1295}
\volume{43}
\issue{3}
\pubyear{2015}
\firstpage{1060}
\lastpage{1088}
\docsubty{FLA}

\makeatletter
\newcommand{\ds}{\displaystyle}
\newcommand{\angler}{\rangle}
\newcommand{\anglel}{\langle}
\newcommand{\eqref}[1]{(\ref{#1})}
\newcommand{\stararrow}{
  \setlength{\unitlength}{1mm}
  \begin{picture}(5,1)(0,0)
    \put(0.2,-0.8){*}
    \put(1,0){$\rightarrow$}
  \end{picture}\,
}

\newcommand{\arrowstar}{
  \setlength{\unitlength}{1mm}
  \begin{picture}(5,1)(0,0)
    \put(0.2,0){$\leftarrow$}
    \put(3,-0.8){*}
  \end{picture}\,
}

\newcommand{\circarrow}{
  \setlength{\unitlength}{1mm}
  \begin{picture}(5,1)(0,0)
    \put(1,1){\circle{1}}
    \put(1,0){$\rightarrow$}
  \end{picture}\,
}
\newcommand{\arrowcirc}{
  \setlength{\unitlength}{1mm}
  \begin{picture}(5,1)(0,0)
    \put(0.2,0){$\leftarrow$}
    \put(4.3,1){\circle{1}}
  \end{picture}\,
}
\newcommand{\tailstar}{
  \setlength{\unitlength}{1mm}
  \begin{picture}(5,1)(0,0)
    \put(0.4,1){\line(1,0){3.2}}
    \put(2.8,-0.8){*}
  \end{picture}\,
}

\newcommand{\startail}{
  \setlength{\unitlength}{1mm}
  \begin{picture}(5,1)(0,0)
    \put(0.2,-1){*}
    \put(1.3,1){\line(1,0){3.2}}
  \end{picture}\,
}

\newcommand{\circstar}{
  \setlength{\unitlength}{1mm}
  \begin{picture}(5,1)(0,0)
    \put(1,1){\circle{1}}
    \put(1.5,1){\line(1,0){2.4}}
    \put(2.9,-0.8){*}
  \end{picture}\,
}
\newcommand{\starcirc}{
  \setlength{\unitlength}{1mm}
  \begin{picture}(5,1)(0,0)
    \put(0.2,-1){*}
    \put(1.1, 1){\line(1,0){2.4}}
    \put(4, 1){\circle{1}}
  \end{picture}\,
}

\newcommand{\circtail}{
  \setlength{\unitlength}{1mm}
  \begin{picture}(5,1)(0,0)
    \put(1,1){\circle{1}}
    \put(1.5,1){\line(1,0){3}}
  \end{picture}
}
\newcommand{\tailtail}{
  \setlength{\unitlength}{1mm}
  \begin{picture}(5,1)(0,0)
    \put(1.5,1){\line(1,0){3}}
  \end{picture}\,
}
\newcommand{\circcirc}{
  \setlength{\unitlength}{1mm}
  \begin{picture}(5,1)(0,0)
    \put(1,1){\circle{1}}
    \put(1.5,1){\line(1,0){2}}
    \put(4,1){\circle{1}}
  \end{picture}\,
}

\newtheorem{lemma}{Lemma}[section]
\newtheorem{theorem}{Theorem}[section]
\newtheorem{corollary}{Corollary}[section]
\newproclaim{remark}{Remark}[section]
\newproclaim{definition}{Definition}[section]
\newproclaim{example}{Example}
\makeatother

\begin{document}
\begin{frontmatter}

\title{A generalized back-door criterion\thanksref{T2}}
\runtitle{A generalized back-door criterion}

\begin{aug}
\author[A]{\fnms{Marloes H.}~\snm{Maathuis}\corref{}\ead[label=e1]{maathuis@stat.math.ethz.ch}}
\and
\author[A]{\fnms{Diego}~\snm{Colombo}\ead[label=e2]{colombo@stat.math.ethz.ch}}
\runauthor{M. H. Maathuis and D. Colombo}
\affiliation{ETH Zurich}
\address[A]{Seminar for Statistics\\
ETH Zurich\\
R\"amistrasse 101\\
8092 Zurich\\
Switzerland\\
\printead{e1}\\
\phantom{E-mail:\ }\printead*{e2}}
\end{aug}
\thankstext{T2}{Supported in part by Swiss NSF Grant 200021-129972.}

%
\received{\smonth{7} \syear{2013}}
\revised{\smonth{11} \syear{2014}}

\begin{abstract}
We generalize Pearl's back-door criterion for directed acyclic graphs
(DAGs) to more general types of graphs that describe Markov equivalence
classes of DAGs and/or allow for arbitrarily many hidden variables. We
also give easily checkable necessary and sufficient graphical criteria
for the existence of a set of variables that satisfies our generalized
back-door criterion, when considering a single intervention and a
single outcome variable. Moreover, if such a set exists, we provide an
explicit set that fulfills the criterion. We illustrate the results in
several examples. R-code is available in the \mbox{R-package} \texttt{pcalg}.
\end{abstract}

\end{frontmatter}

\section{Introduction}

Causal Bayesian networks are widely used for causal reasoning [e.g.,
\citet{GlymourEtAl87,KollerFriedman09}, \citeauthor{Pearl95} (\citeyear{Pearl95,Pearl00,Pearl09book}), \citeauthor{SpirtesEtAl93} (\citeyear{SpirtesEtAl93,SpirtesEtAl00})].
In particular, if the causal structure is known and represented by a
directed acyclic graph (DAG), this framework allows one to deduce
post-intervention distributions and causal effects from the
pre-intervention (or observational) distribution.
%
Hence, if the causal DAG is known, one can estimate causal effects from
observational data. Covariate adjustment is often used for this
purpose. The \textit{back-door criterion} [\citet{Pearl93}] is a
graphical criterion that is sufficient for adjustment, in the sense
that a set of variables can be used for covariate adjustment if it
satisfies the back-door criterion for the given graph.

In practice, there are two important complications. First, the
underlying DAG may be unknown. In this case one can try to estimate the
DAG, but in general one cannot identify the underlying DAG uniquely.
Instead, one can identify its Markov equivalence class, which consists
of all DAGs that encode the same conditional independence relationships
as the underlying DAG. Such a Markov equivalence class can be
represented uniquely by a different type of graph, called a completed
partially directed acyclic graph (CPDAG)
[\citet{SpirtesEtAl93,Meek95,AnderssonEtAl97}]. Second, it is
often the case that some important variables were not measured, meaning
that we do not have causal sufficiency. In this case, one can work with
maximal ancestral graphs (MAGs) instead of DAGs
[\citeauthor{RichardsonSpirtes02} (\citeyear{RichardsonSpirtes02,RichardsonSpirtes03})]. Finally, the
underlying MAG may be unknown, so that it must be estimated from data.
Again, there is an identifiability problem here, as we can generally
only identify the Markov equivalence class of the underlying MAG, which
can be represented uniquely by a partial ancestral graph (PAG)
[\citet{RichardsonSpirtes02,AliRichardsonSpirtes09}].

 In this paper, we therefore consider generalizations of the back-door criterion to the following three scenarios:
       \begin{longlist}[(3)]
          \item[(1)] we assume causal sufficiency, and we only know the CPDAG, that is, the Markov equivalence class of the underlying DAG;
          \item[(2)] we do \textit{not} assume causal sufficiency, and we know the MAG on the observed variables;
          \item[(3)] we do \textit{not} assume causal sufficiency, and we only know the PAG, that is, the Markov equivalence class of the underlying MAG on the observed variables.
       \end{longlist}
In scenarios 2 and 3, we allow for arbitrarily many hidden (or
unmeasured) variables. We do not, however, allow for selection
variables, that is, for unmeasured variables that determine whether a
unit is included in the sample.

Since the back-door criterion is a simple criterion that is widely used
for DAGs, it seems useful to have similar criteria for CPDAGs, MAGs and
PAGs. We also hope that our generalized back-door criterion will make
working with MAGs and PAGs less daunting, and more accessible to people
in practice.

Our generalized back-door criterion for DAGs, CPDAGs, MAGs and PAGs is
given in Section~\ref{secgeneralizedbackdoorcriterion}; see especially
Definition~\ref{defbackdoorgraphical} and Theorem~\ref{thgeneralizedbackdoorsufficientforadjustment}. Corresponding
R-code is available in the function
  \texttt{backdoor} in the R-package \texttt{pcalg} [\citet{KalischEtAl12}]. Our results are derived by first
formulating invariance conditions that are sufficient for adjustment,
and then using the graphical criteria for invariance derived by
\citet{Zhang08-causal-reasoning-ancestral-graphs}. We also show that the
generalized back-door criterion is equivalent to Pearl's back-door
criterion for single interventions in DAGs, and is slightly more
general for multiple interventions in DAGs (Lemma~\ref{lemmaourbackdoorstrongerthanPearls} and  Example~\ref{exourbackdoorstrongerthanPearls}). In Section~\ref{secfindingbackdoorset}, we give necessary and sufficient criteria
for the existence of a set that satisfies the generalized back-door
criterion relative to a pair of variables $(X,Y)$ and a DAG, MAG, CPDAG
or PAG. Moreover, if a generalized back-door set exists, we provide an
explicit such set. These results are summarized in Theorem~\ref{thbackdoorsetforgeneralgraph}, using a general framework that
covers DAGs, CPDAGs, MAGs and PAGs. Corollaries
\ref{corbackdoorsetDAG}--\ref{corbackdoorsetMAG} specialize the results
for DAGs, CPDAGs and MAGs, respectively. We illustrate our results with
several examples in Section~\ref{secexamples}. All proofs are given in
Section~\ref{secproofs}.

We close this introduction by discussing related work. For a given
causal DAG, identifiability of causal effects in general or via
covariate adjustment has been studied by various authors. In
particular, there are complete graphical criteria for the
identification of causal effects if a causal DAG with unmeasured
variables is given [e.g.,
\citet{HuangValtorta06}, \citeauthor{ShpitserPearl06a} (\citeyear{ShpitserPearl06a,ShpitserPearl06b,ShpitserPearl08}), \citet{TianPearl02}].
Shpitser, Van~der Weele and\break Robins (\citeyear{ShpitserVanderWeeleRobins10,ShpitserVanderWeeleRobins10appendum})
studied effects that are identifiable via covariate adjustment, and
provided necessary and sufficient graphical criteria for this purpose,
again if the causal DAG is given. Their results can be viewed as an
improvement on the back-door criterion, which is only sufficient for
adjustment. \citet{TextorLiskiewicz11} studied covariate adjustment for
a given DAG from an algorithmic perspective. Among other things, they
showed that the back-door criterion and the adjustment criterion of
\citet{ShpitserVanderWeeleRobins10} are equivalent if one is
interested in minimal adjustment sets for a certain subclass of graphs.
\citet{VanderZanderEtAl14} extended these necessary and sufficient
graphical criteria for covariate adjustment to MAGs.

There are also existing approaches that do not make the assumption that
the causal DAG or MAG is given.  The prediction algorithm
[\citet{SpirtesEtAl00}, Chapter~7] roughly starts from a PAG and
uses invariance results. In this sense it is probably closest to our
work. The main difference between this method and our results is that
the prediction algorithm is more complex. In particular, it searches
over all possible orderings of the variables, which quickly becomes
infeasible for large graphs. The prediction algorithm may, however, be
more informative, in the sense that certain distributions may be
identifiable by the prediction algorithm but not by the generalized
back-door criterion. Studying the exact relationship between these two
approaches would be an interesting topic for future work.

Other work on data driven methods for selection of adjustment variables
for the estimation of causal effects does not assume that the causal
structure is known, but does make some assumptions about causal
relationships between the variables of interest and/or about the
existence of a set of variables that can be used for covariate
adjustment
[\citet{DeLunaEtAl11,VanderWeeleSphitser11,EntnerHoyerSpirtes13}].
In the current paper, we do not make any such assumptions. On the other
hand, we start from a given DAG, CPDAG, MAG or PAG. We do not see this
as a genuine restriction of our approach, however, since there are
algorithms to estimate CPDAGs and PAGs from data (e.g., the PC
algorithm [\citet{SpirtesEtAl00}], greedy equivalence search
[\citet{Chickering02}] and versions of the FCI algorithm
[\citet{SpirtesEtAl00,ColomboEtAl12,ClaassenMooijHeskes13}]).
These algorithms have been shown to be consistent, even in certain
sparse high-dimensional settings
[\citet{KalischBuehlmann07a,ColomboEtAl12}]. In practice, one
could therefore first employ such an algorithm, and then apply the
results in the current paper.

\section{Preliminaries}

Throughout this paper, we denote sets in a bold font (e.g., $\mathbf X$) and graphs in a calligraphic font (e.g., $\mathcal D$ or $\mathcal M$).

\subsection{Basic graphical definitions}\label{secgraphicaldefinitionsbasic}

A graph $\mathcal{G}=(\mathbf{V},\mathbf{E})$ consists of a set of vertices
$\mathbf{V}=\{X_1,\ldots,X_p\}$ and a set of edges $\mathbf{E}$.
The vertices represent random variables, and the edges
describe conditional independence and causal (ancestral) relationships.
There is at most one edge between
every pair of vertices, and the edge set $\mathbf{E}$ can contain (a subset of) the following four edge types: $\rightarrow$ (\textit{directed}), $\leftrightarrow$
(\textit{bi-directed}), $\circcirc$
(\textit{nondirected}) and
$\circarrow$ (\textit{partially directed}). A \textit{directed graph} contains only directed edges, a \textit{mixed graph} can contain directed and
bi-directed edges and a \textit{partial mixed graph} can contain all four
edge types. The endpoints of an edge
are called \textit{marks}, and they can be \textit{tails},
\textit{arrowheads} or \textit{circles}. We use the symbol ``$*$'' to
denote an arbitrary edge mark. If we are only interested in the presence or absence of edges, and not in the edge marks, then we refer to the \textit{skeleton} of a graph.

Two vertices are \textit{adjacent} if there is an edge between them. The
adjacency set of a vertex $X$ in $\mathcal G$, denoted by
$\operatorname{adj}(X,\mathcal G)$, consists of all vertices adjacent to $X$ in
$\mathcal G$. A \textit{path} is a sequence of distinct adjacent
vertices. The \textit{length
  of a path} $p=\langle X_i, X_{i+1}, \ldots, X_{i+\ell}\rangle$ equals the corresponding number of edges, in this case $\ell$. The path $p$ is said to be
\textit{out of} (\textit{into}) $X_i$ if the edge between $X_i$ and $X_{i+1}$
has a tail (arrowhead) at $X_i$. A sub-path of $p$ from $X_j$ to
$X_{j'}$ is denoted by $p(X_j,X_{j'})$. We denote the concatenation of
paths by $\oplus$, so that, for example, $p = p(X_i,X_{i+k}) \oplus
p(X_{i+k},X_{i+\ell})$ for $k\in \{1,\ldots,\ell-1\}$. We use the
convention that we remove any loops that may occur due to the
concatenation, so that the result does not contain duplicate vertices
and is again a path. The path $p$ is a \textit{directed path} from
$X_i$ to $X_{i+\ell}$ if for all $k\in \{1,\ldots,\ell\}$, the edge
$X_{i+k-1} \to X_{i+k}$ occurs, and it is a \textit{possibly directed
path} if for all $k\in \{1,\ldots,\ell\}$, the edge between $X_{i+k-1}$
and $X_{i+k}$ is not into $X_{i+k-1}$.
  A \textit{cycle} occurs if there is a path between $X_i$ and $X_j$ of length greater than one,
and $X_i$ and $X_j$ are adjacent. A directed path from $X_i$ to $X_j$ forms
a \textit{directed cycle} together with the edge $X_j \rightarrow X_i$, and an \textit{almost directed cycle} together with the edge $X_j \leftrightarrow
X_i$. A \textit{directed acyclic graph} (DAG) is a directed graph without directed cycles. An \textit{ancestral} graph is a mixed graph without directed and almost directed cycles.

If $X_j \rightarrow X_i$, we say that $X_i$ is a \textit{child} of $X_j$,
and $X_j$ is a \textit{parent} of $X_i$. The corresponding sets of parents
and children are denoted by $\operatorname{pa}(X_i, \mathcal{G})$ and $\operatorname{ch}(X_i,
\mathcal{G})$. If there is a (possibly) directed path from $X_i$ to
$X_j$ or if $X_i=X_j$, then $X_i$ is a (\textit{possible}) \textit{ancestor} of
$X_j$ and $X_j$ a (\textit{possible}) \textit{descendant} of $X_i$. The sets of
ancestors, descendants, possible ancestors, and possible descendants of
a vertex $X_i$ in $\mathcal{G}$ are denoted by an$(X_i, \mathcal{G})$,
$\operatorname{de}(X_i,\mathcal{G})$, $\operatorname{possibleAn}(X_i,\mathcal{G})$, and
$\operatorname{possibleDe}(X_i, \mathcal{G})$, respectively. These definitions are
applied disjunctively to a set $\mathbf{Y} \subseteq \mathbf{V}$, for
example, $\operatorname{an}(\mathbf{Y}, \mathcal{G}) = \{X_i    \vert    X_i
\in \operatorname{an} (X_j, \mathcal{G}) \mbox{ for some } X_j \in
\mathbf{Y}\}$.

A path $\langle X_i,X_j,X_k \rangle$ is an \textit{unshielded triple}
if $X_i$ and $X_k$ are not adjacent. A~nonendpoint vertex $X_j$ on a path is
a \textit{collider} on the path if the path contains~$\stararrow X_j
\arrowstar\!$. A nonendpoint vertex on a path which is not a
collider is a \textit{noncollider} on the path.
 A \textit{collider
  path} is a path on which every nonendpoint vertex is a
collider. A path of length one is a trivial collider path.

%

\subsection{Causal Bayesian networks}

A Bayesian network for a set of variables
$\mathbf{V}=\{X_1,\ldots,X_p\}$ is a pair $(\mathcal D, f)$, where
$\mathcal D = (\mathbf{V},\mathbf{E})$ is a DAG, and $f$ is a joint
probability density for $\mathbf{V}$ (with respect to some dominating
measure) that factorizes according to $\mathcal D$: $f(\mathbf{V}) =
\prod_{i=1}^p f(X_i|\operatorname{pa}(X_i, \mathcal D))$. If the DAG is
interpreted causally, in the sense that $X_i \to X_j$ means that $X_i$
has a (potential) direct causal effect on $X_j$, then we talk about a
\textit{causal DAG} and a \textit{causal Bayesian network}.

One can easily derive post-intervention densities if the causal
Bayesian network is given and all variables are observed. In
particular, we consider interventions $\operatorname{do}(\mathbf{X} = \mathbf{x})$ for
$\mathbf{X}\subseteq \mathbf{V}$ [\citet{Pearl00}], which
represent outside interventions that set the variables in $\mathbf{X}$
to their respective values in $\mathbf{x}$. We assume that such
interventions are effective, meaning that $\mathbf{X}=\mathbf{x}$ after
the intervention. Moreover, we assume that the interventions are local,
meaning that the generating mechanisms of the other variables, and
hence their conditional distributions given their parents, do not
change. We then have
\begin{eqnarray*}
&& f\bigl(\mathbf{V}|\operatorname{do}(\mathbf{X} = \mathbf{x})\bigr)
\\
&&\qquad = \cases{\ds\prod_{X_i\in \mathbf{V}\setminus \mathbf{X}} f
\bigl(X_i|\operatorname{pa}(X_i,\mathcal D)\bigr), & \quad \mbox{for values of $\mathbf{V}$ consistent with $\mathbf{x}$},
\vspace*{3pt}\cr
0,  & \quad \mbox{otherwise}.}
\end{eqnarray*}
This is known as the g-formula or the truncated factorization formula
[\citet{Robins86,SpirtesEtAl93,Pearl00}].

In a Bayesian network $(\mathcal D, f)$, the DAG $\mathcal D$ encodes
conditional independence relationships in the density $f$ via
d-separation [\citet{Pearl00}; see also Definition~\ref{defm-connection}]. Several DAGs can encode the same conditional
independence relationships. Such DAGs form a Markov equivalence class
which can be uniquely represented by a CPDAG. A CPDAG is a graph with
the same skeleton as each DAG in its equivalence class, and its edges
are either directed ($\to$) or nondirected ($\circcirc\!$). An edge $X_i
\to X_j$ in such a CPDAG means that $X_i \to X_j$ is present in every
DAG in the Markov equivalence class, while an edge $X_i \circcirc X_j$
represents uncertainty about the edge marks, in the sense that the
Markov equivalence class contains at least one DAG with $X_i\to X_j$
and at least one DAG with $X_i \leftarrow X_j$. (Note that many authors
use $X_i \tailtail X_j$ instead of $X_i \circcirc X_j$; we use
$\circcirc$ to ensure that the CPDAG satisfies the syntactic properties
of a PAG; see below.)

If some of the variables in a DAG are unobserved, one can transform the
DAG into a unique \textit{maximal ancestral graph} (MAG) on the observed
variables; see \citeauthor{RichardsonSpirtes02}
[(\citeyear{RichardsonSpirtes02}), page 981] for an algorithm.
In particular, two vertices $X_i$ and $X_j$ are adjacent in a MAG if
and only if no subset of the remaining observed variables makes them
conditionally independent. Moreover, a tail mark $X_i \tailstar X_j$ in
a MAG $\mathcal M$ means that $X_i$ is an ancestor of $X_j$ in all DAGs
represented by $\mathcal M$, while an arrowhead $X_i \arrowstar X_j$
means that $X_i$ is not an ancestor of $X_j$ in all DAGs represented by
$\mathcal M$. Thus an edge $X_i \to X_j$ in $\mathcal M$ means that
there is a directed path from $X_i$ to $X_j$ in all DAGs represented by
$\mathcal M$, but we emphasize that it does not represent a direct
effect with respect to the observed variables, in the sense that there
may be other observed variables on the directed path. Several different
DAGs can lead to the same MAG, and a MAG represents a class of
(infinitely many) DAGs that have the same d-separation and ancestral
relationships among the observed variables. The MAG of a causal DAG is
called a \textit{causal MAG}.

A MAG encodes conditional independence relationships via the concept of
m-separation (Definition~\ref{defm-connection}). Again, several MAGs
can encode the same conditional independence relationships. Such MAGs
are called Markov equivalent, and can be uniquely represented by a
\textit{partial ancestral graph} (PAG). This is a partial mixed graph
with the same skeleton as each MAG in its Markov equivalence class. A~tail mark (arrowhead) at an edge $X_i \tailstar X_j$ ($X_i \arrowstar
X_j)$ in such a PAG means that $X_i \tailstar X_j$ ($X_i \arrowstar
X_j$) in every MAG in the Markov equivalence class, while a circle mark
at an edge $X_i \circstar X_j$ represents uncertainty about the edge
mark, in the sense that the Markov equivalence class contains at least
one MAG with $X_i \tailstar X_j$, and at least one MAG with $X_i
\arrowstar X_j$.


We say that a density $f$ is \textit{compatible with a DAG $\mathcal D$}
if the pair $(\mathcal D, f)$ forms a causal Bayesian network. A
density $f$ is \textit{compatible with a CPDAG $\mathcal C$} if it is
compatible with a DAG in the Markov equivalence class described by
$\mathcal C$. A density $f$ is \textit{compatible with a MAG $\mathcal
M$} if there exists a causal Bayesian network $(\mathcal D^*, f^*)$
(including hidden variables), such that $\mathcal M$ is the MAG of
$\mathcal D^*$ and $f$ is the corresponding marginal of $f^*$.
Finally, $f$ is \textit{compatible with a PAG $\mathcal P$} if it is
compatible with a MAG in the Markov equivalence class described by
$\mathcal P$.

\section{Generalized back-door criterion}\label{secgeneralizedbackdoorcriterion}

We now present our generalized back-door criterion in Definition~\ref{defbackdoorgraphical} and Theorem~\ref{thgeneralizedbackdoorsufficientforadjustment}, where the name
``generalized back-door criterion'' is motivated by Lemma~\ref{lemmaourbackdoorstrongerthanPearls}. We first introduce some more
specialized definitions.

\citet{Zhang08-causal-reasoning-ancestral-graphs} introduced the concept
of (\textit{definitely}) \textit{visible} edges in MAGs and PAGs. The reason for
this is as follows. A directed edge $X\to Y$ in a DAG, CPDAG, MAG or
PAG always means that $X$ is a cause (or ancestor) of $Y$, because of
the tail mark at $X$. However, if we allow for hidden variables (i.e.,
in MAGs and PAGs), there may be a hidden confounding variable between
$X$ and $Y$. Visible edges refer to situations where there cannot be
such a hidden confounder between $X$ and $Y$. Invisible edges, on the
other hand, are possibly confounded in the sense that there is a DAG
represented by the MAG or PAG with $X \leftarrow L \to Y$, where $L$ is
not measured (in addition to $X\to \cdots \to Y$).

\begin{definition}[{[Visible and invisible edges; cf. \citet{Zhang08-causal-reasoning-ancestral-graphs}]}]\label{defvisibleedge}
   All directed edges in DAGs and CPDAGs are said to be \textit{visible}. 
   Given a $\operatorname{MAG} \mathcal M /\break  \operatorname{PAG} \mathcal P$, a directed edge $A \to B$ in $
   \mathcal M/\mathcal P$ is \textit{visible} if there is a vertex $C$ not adjacent to $B$, such that there is an edge between $C$ and $A$ that is into $A$, or there is a
collider path between $C$ and $A$ that is into $A$ and every nonendpoint vertex on the path is a parent of $B$. Otherwise
$A \to B$ is said to be invisible.
\end{definition}

\begin{figure}

\includegraphics{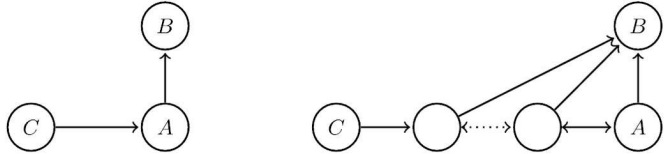}

  \caption{Edge configurations in MAGs and PAGs for a visible
    edge $A \to B$; cf. \protect\citet{Zhang08-causal-reasoning-ancestral-graphs}, Figure~6.
   Instead of the tail mark at $C$, one can also have an arrowhead or circle mark.}
  \label{figvisibileedges}
\end{figure}

Figure~\ref{figvisibileedges} illustrates the different graphical
configurations that can lead to a visible edge. We note that
\citet{Zhang08-causal-reasoning-ancestral-graphs} used slightly
different terminology, referring to \textit{definitely visible} edges in
a PAG, while we simply say \textit{visible} for both MAGs and PAGs.
\citet{BorboudakisTriantafillouTsamardinos12} used the term
\textit{pure-causal} edges instead of \textit{visible} edges in MAGs.

We can now generalize the concept of a \textit{back-door path} in
Definition~\ref{defbackdoorpath}.

\begin{definition}[(Back-door path)]\label{defbackdoorpath}
Let $(X,Y)$ be an ordered pair of vertices in $\mathcal{G}$, where $\mathcal{G}$
is a DAG, CPDAG, MAG or PAG. We say that a path between $X$ and $Y$ is
a \textit{back-door path from $X$ to $Y$} if it does not have a visible
edge out of $X$.
\end{definition}

In a DAG, this definition reduces to a path between $X$ and $Y$ that
starts with $X\leftarrow$, which is the usual back-door path as defined
by \citet{Pearl93}. In a CPDAG, a back-door path from $X$ to $Y$ is a
path between $X$ and $Y$ that starts with $X \leftarrow$ or $X
\circcirc\!$. In a MAG, it is a path between $X$ and $Y$ that starts with
$X \leftrightarrow$, $X\leftarrow$ or an invisible edge $X
\rightarrow$. Finally, in a PAG, it is a path between $X$ and $Y$ that
starts with $X \arrowstar\!$, $X \circstar $ or an invisible edge $X
\rightarrow$.

We also need generalizations of the concept of
d-separation in DAGs [Definition~1.2.3 of \citet{Pearl00}]. In MAGs, one can use m-separation
[Section~3.4 of \citet{RichardsonSpirtes02}]. In CPDAGs and PAGs, there
is the additional complication that it may be unclear whether a vertex
is a collider or a noncollider on the path. We therefore need the
following definitions:

\begin{definition}[{[Definite noncollider; \citet{Zhang08-causal-reasoning-ancestral-graphs}]}]\label{defdefinitenoncollider}
A nonendpoint vertex $X_j$ on a path $\anglel \ldots, X_i,X_j,X_k, \ldots\angler $ in a partial mixed graph $\mathcal G$ is a
\textit{definite noncollider} on the path if (i) there is a tail mark at
$X_j$, that is, $X_i \startail X_j$ or $X_j \tailstar X_k$, or (ii)
$\anglel X_i,X_j,X_k\angler $ is unshielded and has circle marks at $X_j$,
that is, $X_i \starcirc X_j \circstar X_k$ and $X_i$ and $X_k$ are not
adjacent in $\mathcal G$.
\end{definition}

The motivation for conditions (i) and (ii) is straightforward. A tail
mark out of~$X_j$ on the path ensures that $X_j$ is a noncollider on
the path in any graph obtained by orienting any possible circle marks.
Condition (ii) comes from the fact that the collider status of
unshielded triples is known in CPDAGs and PAGs. Hence, if the graph
contains an unshielded triple that was not oriented as a collider, then
it must be a noncollider in all underlying DAGs or MAGs. If $\mathcal
G$ is a DAG or a MAG, then only condition (i) applies and reduces to
the usual definition of a noncollider.

\begin{definition}[(Definite status path)]
  A nonendpoint vertex $X_j$ on a path $p$ in a partial mixed graph is said to be of a \textit{definite status} if
  it is either a collider or a definite noncollider on $p$. The path $p$ is said to be of a \textit{definite status}
  if all nonendpoint vertices on the path are of a definite status.
\end{definition}

A path of length one is a trivial definite status path. Moreover, in DAGs and MAGs, all paths are of a definite status.

We now define m-connection for definite status paths.

\begin{definition}[(m-connection)]\label{defm-connection}
   A definite status path $p$ between vertices $X$ and $Y$ in a partial mixed graph is
    \textit{m-connecting} given a (possibly empty) set of variables
    $\mathbf{Z}$ $(X,Y \notin\mathbf{Z})$ if the following two conditions hold:
    \begin{longlist}[(a)]
       \item [{(a)}] every definite noncollider on the path is not in $\mathbf{Z}$;
       \item [{(b)}] every collider on the path is an ancestor of some
         member of $\mathbf{Z}$.
    \end{longlist}
    If a definite status path $p$ is not m-connecting given $\mathbf{Z}$, then we say that $\mathbf{Z}$ blocks~$p$.
\end{definition}

If $\mathbf{Z}=\varnothing$, we usually omit the phrase ``given the
empty set.'' Definition~\ref{defm-connection} reduces to m-connection
for MAGs and d-connection for DAGs. We note that
\citet{Zhang08-causal-reasoning-ancestral-graphs} used the notions of
\textit{possible m-connection} and \textit{definite m-connection} in PAGs,
where his notion of definite m-connection is the same as our notion of
m-connection for definite status paths.

We now define an adjustment criterion for DAGs, CPDAGs, MAGs and PAGs. Throughout, we think of $\mathbf{X}$ and $\mathbf{Y}$ as nonempty sets.

\begin{definition}[(Adjustment criterion)]\label{defadjustmentcriterion}
   Let $\mathbf{X}$, $\mathbf{Y}$ and $\mathbf{W}$ be pairwise disjoint sets of
   vertices in $\mathcal{G}$, where $\mathcal{G}$ represents a DAG, CPDAG, MAG or PAG. Then we say that $\mathbf{W}$ satisfies the adjustment criterion
   relative to $(\mathbf{X},\mathbf{Y})$ and $\mathcal G$ if for any density $f$ compatible with $\mathcal G$, we have
\begin{eqnarray*}
f\bigl(\mathbf{y}|\operatorname{do}(\mathbf{x})\bigr) = \cases{f(\mathbf{y}|
\mathbf{x}),  &\quad  \mbox{if $\mathbf{W}=\varnothing$},
\vspace*{3pt}\cr
\ds \int_{\mathbf{w}} f(\mathbf{y}|\mathbf{w},\mathbf{x}) f(
\mathbf{w})\,d\mathbf{w} = E_{\mathbf{W}} \bigl\{ f(\mathbf{y}|\mathbf{w},
\mathbf{x})\bigr\}, &\quad \mbox{otherwise}.}
\end{eqnarray*}
\end{definition}

If $\mathbf{X}=\{X\}$ and $\mathbf{Y}=\{Y\}$, we simply say that a set satisfies the criterion relative to $(X,Y)$ [rather than $(\{X\},\{Y\})$] and the given graph.


We now propose our generalized back-door criterion for DAGs, CPDAGs,
MAGs and PAGs. We will show in Theorem~\ref{thgeneralizedbackdoorsufficientforadjustment} that this criterion
is sufficient for adjustment.

\begin{definition}[(Generalized back-door criterion and generalized back-door set)]\label{defbackdoorgraphical}
   Let $\mathbf{X}$, $\mathbf{Y}$ and $\mathbf{W}$ be pairwise disjoint sets of
   vertices in $\mathcal{G}$, where $\mathcal{G}$ represents a DAG, CPDAG, MAG or PAG. Then $\mathbf{W}$ satisfies the \textit{generalized back-door criterion} relative to
   $(\mathbf{X},\mathbf{Y})$ and $\mathcal G$ if the following two conditions hold:
  \begin{longlist}[(B-ii)]
     \item[{(B-i)}] $\mathbf{W}$ does not contain possible descendants of $\mathbf{X}$ in $\mathcal G$;
  \item[{(B-ii)}] for every $X \in \mathbf{X}$, the set $\mathbf{W}
      \cup \mathbf{X} \setminus \{X\}$ blocks every definite status
      back-door path from $X$ to any member of $\mathbf{Y}$, if
      any, in $\mathcal G$.
  \end{longlist}
  A set $\mathbf{W}$ that satisfies the generalized back-door criterion relative to $(\mathbf{X},\mathbf{Y})$ and $\mathcal G$ is called a
  \textit{generalized back-door set} relative to $(\mathbf{X},\mathbf{Y})$ and $\mathcal G$.
\end{definition}

\begin{remark}\label{rempossibledescendantsdefinitestatuspath}
  Condition (B-i) in Definition~\ref{defbackdoorgraphical} is equivalent to the following:
  \begin{longlist}[(B-i)$'$]
     \item[{(B-i)$'$}] $\mathbf{W}$ does not contain possible descendants of $\mathbf{X}$ \textit{along a definite status path} in $\mathcal G$.
  \end{longlist}
Condition (B-i)$'$ may be easier to check computationally than (B-i). The
equivalence of (B-i) and (B-i)$'$ is shown in the proof of Theorem~\ref{thgeneralizedbackdoorsufficientforadjustment}, using Lemma~\ref{lemmapossibledescendantsdefinitestatuspath}.
\end{remark}

\begin{theorem}\label{thgeneralizedbackdoorsufficientforadjustment}
   Let $\mathbf{X}$, $\mathbf{Y}$ and $\mathbf{W}$ be pairwise disjoint sets of
   vertices in $\mathcal{G}$, where $\mathcal{G}$ represents a DAG, MAG, CPDAG or PAG. If $\mathbf{W}$ satisfies the \textit{generalized back-door criterion} relative to
$(\mathbf{X},\mathbf{Y})$ and $\mathcal G$ (Definition~\ref{defbackdoorgraphical}), then it satisfies the adjustment criterion
relative to $(\mathbf{X},\mathbf{Y})$ and $\mathcal G$ (Definition~\ref{defadjustmentcriterion}).
\end{theorem}

The proof of Theorem~\ref{thgeneralizedbackdoorsufficientforadjustment}
consists of two steps. First, we formulate invariance criteria that are
sufficient for adjustment (Theorem~\ref{thinvarianceimpliesadjustment}). Next, we translate the invariance
criteria into the graphical criteria given in Definition~\ref{defbackdoorgraphical}, using results of
\citet{Zhang08-causal-reasoning-ancestral-graphs} (Theorem~\ref{thbackdoorequivalenttoinvariance}).

We refer to Definition~\ref{defbackdoorgraphical} as generalized
back-door criterion because its conditions are closely related to
Pearl's original back-door criterion [\citeauthor{Pearl93} (\citeyear{Pearl93,Pearl00})].

\begin{definition}[{[Pearl's back-door criterion; Definition~3.3.1 of \citet{Pearl00}]}]\label{defbackdoorPearl}
   A set of variables $\mathbf{W}$ satisfies the back-door criterion relative to an ordered pair of variables $(X,Y)$ in a DAG $\mathcal D$
   if the following two conditions hold:
  \begin{longlist}[(P-ii)]
     \item[{(P-i)}] no vertex in $\mathbf{W}$ is a descendant of $X$ in $\mathcal D$;
     \item[{(P-ii)}] $\mathbf{W}$ blocks every path between $X$ and $Y$ in $\mathcal D$ that is into $X$.
  \end{longlist}
Similarly, if $\mathbf{X}$ and $\mathbf{Y}$ are two disjoint subsets of
vertices in $\mathcal D$, then $\mathbf{W}$ is said to satisfy the
back-door criterion relative to $(\mathbf{X},\mathbf{Y})$ in $\mathcal
D$ if it satisfies the criterion relative to any pair $(X,Y)$ such that
$X\in \mathbf{X}$ and $Y\in \mathbf{Y}$.
\end{definition}

In particular, the conditions in Definition~\ref{defbackdoorgraphical}
are equivalent to Pearl's back-door criterion for a DAG with a single
intervention ($|\mathbf{X}|=1$). For a DAG with multiple interventions,
any set that satisfies Pearl's back-door criterion also satisfies the
generalized back-door criterion, but not necessarily the other way
around. In this sense, our criterion is slightly better; see Lemma~\ref{lemmaourbackdoorstrongerthanPearls} and Example~\ref{exourbackdoorstrongerthanPearls}.

\begin{lemma}\label{lemmaourbackdoorstrongerthanPearls}
Let $\mathbf{X}$, $\mathbf{Y}$ and $\mathbf{W}$ be pairwise disjoint
sets of vertices in a DAG $\mathcal D$. If $\mathbf{W}$ satisfies
Pearl's back-door criterion (Definition~\ref{defbackdoorPearl})
relative to $(\mathbf{X},\mathbf{Y})$ and~$\mathcal D$, then
$\mathbf{W}$ satisfies the generalized back-door criterion (Definition~\ref{defbackdoorgraphical}) relative to $(\mathbf{X},\mathbf{Y})$ and
$\mathcal D$.
\end{lemma}

\section{Finding a set that satisfies the generalized back-door criterion}\label{secfindingbackdoorset}

An important reason for the popularity of Pearl's back-door criterion
is the following. Consider two distinct vertices $X$ and $Y$ in a DAG
$\mathcal D$. Then $\operatorname{pa}(X,\mathcal D)$ satisfies the back-door
criterion relative to $(X,Y)$ and $\mathcal D$, unless $Y\in
\operatorname{pa}(X,\mathcal D)$. In the latter case, there is no set that
satisfies the back-door criterion relative to $(X,Y)$ and $\mathcal D$,
but it is easy to see that $f(y|\operatorname{do}(x))=f(y)$ for any density $f$
compatible with $\mathcal D$, since there cannot be a directed path
from $X$ to $Y$ in $\mathcal D$.

In this section, we formulate similar results for the generalized
back-door criterion. In particular, we consider the following problem.
Given two distinct vertices $X$ and $Y$ in a DAG, CPDAG, MAG or PAG,
can we easily determine if there exists a generalized back-door set
relative to $(X,Y)$ and the given graph? Moreover, if this question is
answered positively, can we give an explicit set that satisfies the
criterion? Theorem~\ref{thbackdoorsetforgeneralgraph} addresses these
questions in general, while Corollaries
\ref{corbackdoorsetDAG}--\ref{corbackdoorsetMAG} give specific results
for DAGs, CPDAGs and MAGs.

We emphasize that throughout this section, we focus on the setting with
a single intervention variable $X$ and a single variable of interest
$Y$. The setting with multiple interventions (i.e., a set $\mathbf{X}$)
is considerably more difficult, even for DAGs
[\citet{ShpitserVanderWeeleRobins10}]. It therefore seems
challenging to generalize the results in this section to sets
$\mathbf{X}$. Handling sets $\mathbf{Y}$ seems less difficult, and we
plan to study this in future work.

In a DAG, the following result is well known. If $X$ and $Y$ are not
adjacent in a DAG $\mathcal D$ and $X\notin \operatorname{an}(Y,\mathcal D)$,
then $\operatorname{pa}(X,\mathcal D)$ blocks all paths between $X$ and $Y$. In
MAGs, we have a similar result, but we need to use
$\operatorname{D\mbox{-}SEP}(X,Y,\mathcal M)$ instead of the parent set; see
Definition~\ref{defD-SEP} and Lemma~\ref{lemmaD-SEP}.

\begin{definition}[{[D-SEP$(X,Y,\mathcal G)$; cf. page 136 of \citet{SpirtesEtAl00}]}]\label{defD-SEP}
  Let $X$ and $Y$ be two distinct vertices in a mixed graph $\mathcal{G}$. We say
  that $V \in \operatorname{D\mbox{-}SEP}(X,Y,\mathcal G)$ if $V \neq X$, and there is a
  collider path between $X$ and $V$ in $\mathcal G$, such that every vertex
  on this path (including $V$) is an ancestor of $X$ or $Y$ in $\mathcal G$.
\end{definition}

\begin{lemma}\label{lemmaD-SEP}
  Let $X$ and $Y$ be two distinct vertices in an ancestral graph $\mathcal G$. Then the following statements are equivalent:
\textup{(i)} $X$ and $Y$ are m-separated in $\mathcal G$ by some subset of the
remaining variables, \textup{(ii)} $Y\notin \operatorname{D\mbox{-}SEP}(X,Y,\mathcal G)$, and
\textup{(iii)} $X$ and $Y$ are m-separated in $\mathcal G$ by
$\operatorname{D\mbox{-}SEP}(X,Y,\mathcal G)$.
  Moreover, if $\mathcal G$ is a MAG, a fourth equivalent statement is \textup{(iv)} $X$ and $Y$ are not adjacent in $\mathcal G$.
\end{lemma}

We now introduce important definitions that are needed to formulate our
generalized back-door criterion in Theorem~\ref{thbackdoorsetforgeneralgraph}.

\begin{definition}[($\mathcal R^*$ and $\mathcal R_{\underline X}$)] \label{defR}
   Let $X$ be a vertex in $\mathcal{G}$, where $\mathcal{G}$ is a DAG, CPDAG, MAG or PAG.

Let $\mathcal R^* = \mathcal R^*(\mathcal G,X)$ be a class of DAGs or
MAGs, defined as follows. If $\mathcal G$ is a DAG or a MAG, we simply
let $\mathcal R^* = \{\mathcal G\}$. If $\mathcal G$ is a CPDAG/PAG,
we let $\mathcal R^*$ be the subclass of DAGs/MAGs in the Markov
equivalence class described by $\mathcal G$ that have the same number
of edges into $X$ as $\mathcal G$.

For any $\mathcal R \in \mathcal R^*$, let $\mathcal R_{\underline
X}=\mathcal R_{\underline X}(\mathcal R, \mathcal G, X)$ be the graph
obtained from $\mathcal R$ by removing all directed edges out of $X$
that are visible in $\mathcal G$; see Definition~\ref{defvisibleedge}.
\end{definition}

For any given $\mathcal G$ and $X$, we say that a graph $\mathcal
R_{\underline X}$ satisfies Definition~\ref{defR} if there exists an
$\mathcal R \in \mathcal R^*(\mathcal G, X)$ such that $\mathcal
R_{\underline X} = \mathcal R_{\underline X}(\mathcal R, \mathcal G,
X)$.

Lemma~\ref{lemmaMAGwithsameindegreeasPAG} shows that the class
$\mathcal R^*$ is always nonempty. The definition of $\mathcal
R_{\underline X}$ is related to the $X$-lower manipulated MAGs that
were used by \citet{Zhang08-causal-reasoning-ancestral-graphs}. It is
important to note, however, that $\mathcal R_{\underline X}$ is
obtained from $\mathcal R$ by removing the edges out of $X$ that are
visible in $\mathcal G$ (rather than $\mathcal R$). Moreover, Zhang
replaced invisible edges by bi-directed edges, but that is not needed
for our purposes (although it would not hurt to do so). Finally, we
note that $\mathcal R_{\underline X}$ is ancestral, since any $\mathcal
R\in \mathcal R^*$ is ancestral.

We can now present the main result of this section.

\begin{theorem}[(Generalized back-door set)]\label{thbackdoorsetforgeneralgraph}
Let $X$ and $Y$ be two distinct vertices in $\mathcal{G}$, where $\mathcal{G}$ is
a DAG, CPDAG, MAG or PAG. Let $\mathcal R_{\underline X}$ be any graph
satisfying Definition~\ref{defR}. Then there exists a generalized
back-door set relative to $(X,Y)$ and $\mathcal{G}$ if and only if $Y\notin
\operatorname{adj}(X,\mathcal R_{\underline X})$ and
   $\operatorname{D\mbox{-}SEP}(X,Y,\break \mathcal R_{\underline X}) \cap
   \operatorname{possibleDe}(X,\mathcal G) = \varnothing$. Moreover, if such a generalized back-door set exists, then $\operatorname{D\mbox{-}SEP}(X,Y,\mathcal R_{\underline X})$ is such a set.
\end{theorem}

The definitions of $\mathcal R^*$ and $\mathcal R_{\underline X}$ in
Definition~\ref{defR} are needed in Theorem~\ref{thbackdoorsetforgeneralgraph} to ensure that
$\operatorname{D\mbox{-}SEP}(X,Y,\mathcal R_{\underline X}) \cap
\operatorname{possibleDe}(X,\mathcal G) \neq \varnothing$ implies that there
does not exist a generalized back-door set relative to $(X,Y)$ and
$\mathcal{G}$; see also Example~\ref{exPAG}.

For DAGs, CPDAGs and MAGs we can simplify Theorem~\ref{thbackdoorsetforgeneralgraph} somewhat; see Corollaries
\ref{corbackdoorsetDAG}--\ref{corbackdoorsetMAG}. Corollary~\ref{corbackdoorsetDAG} is the well-known result for DAGs that we
discussed earlier. Corollary~\ref{corbackdoorsetMAG} is given without
proof, since it follows straightforwardly from Theorem~\ref{thbackdoorsetforgeneralgraph}.

\begin{corollary}[(Generalized back-door set for a DAG)]\label{corbackdoorsetDAG}
Let $X$ and $Y$ be two distinct vertices in a DAG $\mathcal D$. Then
there exists a generalized back-door set relative to $(X,Y)$ and $\mathcal{D}$ if and only if $Y\notin \operatorname{pa}(X,\mathcal D)$. Moreover, if such
a generalized back-door set exists, then $\operatorname{pa}(X,\mathcal D)$ is
such a set.
\end{corollary}

\begin{corollary}[(Generalized back-door set for a CPDAG)]\label{corbackdoorsetCPDAG}
Let $X$ and $Y$ be two distinct vertices in a CPDAG $\mathcal C$. Let
$\mathcal C_{\underline X}$ be the graph obtained from $\mathcal C$ by
removing all directed edges out of $X$. Then there exists a generalized
back-door set relative to $(X,Y)$ and $\mathcal{C}$ if and only if $Y\notin
\operatorname{pa}(X,\mathcal C)$ and $Y\notin \operatorname{possibleDe}(X,\mathcal
C_{\underline X})$. Moreover, if such a generalized back-door set
exists, then $\operatorname{pa}(X, \mathcal C)$ is such a set.
\end{corollary}

\begin{corollary}[(Generalized back-door set for a MAG)]\label{corbackdoorsetMAG}
  Let $X$ and $Y$ be two distinct vertices in a MAG $\mathcal M$. Then there exists generalized backdoor set relative to $(X,Y)$ and $\mathcal M$ if and only if
   $Y \notin
  \operatorname{adj}(X,\mathcal{M}_{\underline X})$ and
  $\operatorname{D\mbox{-}SEP}(X,Y,M_{\underline X}) \cap \operatorname{de}(X,\mathcal M) =
  \varnothing$. Moreover, if such a generalized back-door set exists, then $\operatorname{D\mbox{-}SEP}(X,Y,\mathcal{M}_{\underline X})$ is such a set.
\end{corollary}

\section{Examples}\label{secexamples}

We now give several examples to illustrate the theory for DAGs, CPDAGs, MAGs and PAGs.

\subsection{DAG examples}

We start with an example that shows that the generalized back-door criterion for DAGs is weaker than
Pearl's back-door criterion for DAGs, in the sense that it can happen that there is no set that satisfies Pearl's back-door criterion, while there is a set that satisfies the
generalized back-door criterion.

\begin{figure}[b]
\begin{tabular}{@{}cc@{}}

\includegraphics{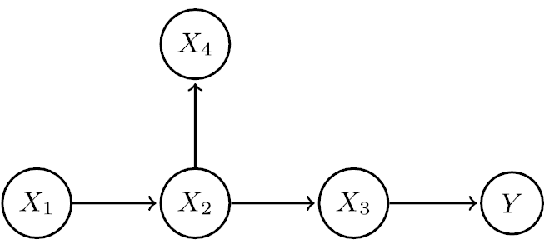}
 & \includegraphics{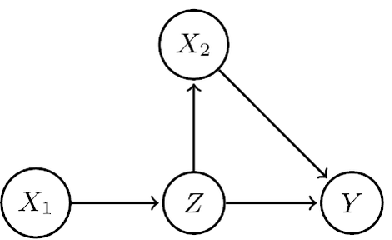}\\
\footnotesize{(a)} & \footnotesize{(b)}
\end{tabular}
\caption{DAG examples. \textup{(a)} The DAG $\mathcal D$ for Example~\protect\ref{exourbackdoorstrongerthanPearls}.
\textup{(b)} The DAG $\mathcal D$ for Example~\protect\ref{exmultintervention}.}\label{figmultintervention}\label{figourbackdoorstrongerthanPearls}
\end{figure}

\begin{example}\label{exourbackdoorstrongerthanPearls}
   Consider the DAG $\mathcal D$ in Figure~\ref{figourbackdoorstrongerthanPearls}(a) with $\mathbf{X} = \{X_1,X_3,X_4\}$ and $\mathbf{Y}
   =\{Y\}$. We first show that $\mathbf{W}=\varnothing$ is a
   generalized back-door set relative to $(\mathbf{X},\mathbf{Y})$ and $\mathcal
   D$. Note that we cannot use Theorem~\ref{thbackdoorsetforgeneralgraph} since $\mathbf{X}$ is a set. We therefore work with Definition~\ref{defbackdoorgraphical} directly. We only need to check that the
   back-door path from $X_4$ to $Y$ is blocked by $\mathbf{W} \cup
   \mathbf{X} \setminus \{X_4\} = \{X_1,X_3\}$, which is the case since
   $X_3$ is a noncollider on the path. Indeed, we have that
   $f(y|\operatorname{do}(x_1,x_3,x_4)) = f(y|x_1,x_3,x_4)$ in Figure~\ref{figourbackdoorstrongerthanPearls}(a), which can be further simplified to
   $f(y|x_3)$.

   On the other hand, there is no set that satisfies Pearl's back-door
   criterion (Definition~\ref{defbackdoorPearl}) with respect to
   $(\mathbf{X},\mathbf{Y})$. To see this, note that
   $\{X_2,X_3,X_4\}\subseteq \operatorname{de}( X_1, \mathcal D)$. Hence, the only
   possible candidate set is $\mathbf{W}=\varnothing$. But this set does not
   block the back-door path from $X_4$ to $Y$, since there is no collider on
   this path.
\end{example}

Next, we note that the generalized back-door criterion is not necessary for identifying post-intervention distributions. Two simple examples are given below.

\begin{example}\label{exxarrowstary}
   Let $X$ and $Y$ be two distinct vertices in $\mathcal G$, where
   $\mathcal G$ represents a DAG, CPDAG, MAG or PAG. If $X
\arrowstar Y$ in $\mathcal G$, then $Y\in \operatorname{adj}(X,\mathcal
R_{\underline X})$ for any $\mathcal R_{\underline X}$ satisfying
Definition~\ref{defR}. Hence Theorem~\ref{thbackdoorsetforgeneralgraph}
implies that there does not exist a generalized back-door set relative
to $(X,Y)$ and $\mathcal G$.

   On the other hand, it is clear that $f(y|\operatorname{do}(x)) = f(y)$ for any density $f$ compatible with $\mathcal G$, since
   the edge $X \arrowstar Y$ implies that there cannot be a possibly directed path from $X$ to
   $Y$ in $\mathcal G$; see Lemma~\ref{lemmapossiblydirectedpathandedgeinto} below.
\end{example}

\begin{example}\label{exmultintervention}
   Let $\mathcal D$ be the DAG in Figure~\ref{figmultintervention}(b), and let $\mathbf{X} = \{X_1,X_2\}$ and $\mathbf{Y}=\{Y\}$.
   Then there does not exist a generalized
   back-door set relative to
$(\mathbf{X},\mathbf{Y})$ and~$\mathcal D$. To see this, note that the
only candidate variable $Z$ cannot be used, since it is a descendant of
$X_1$. Moreover, $\mathbf{W} = \varnothing$ violates condition (B-ii)
in Definition~\ref{defbackdoorgraphical} for~$X_2$, since $\mathbf{W}
\cup \mathbf{X} \setminus \{X_2\} = \{X_1\}$ does not block the
back-door path \mbox{$X_2 \leftarrow Z \rightarrow Y$}.

   On the other hand, $f(y|\operatorname{do}(x_1,x_2)) = \int f(z|x_1)f(y|x_2,z)\,dz$ for any density $f$ compatible with $\mathcal D$, by the g-formula.
\end{example}

\subsection{CPDAG examples}

\begin{figure}[b]
\begin{tabular}{@{}cc@{}}

\includegraphics{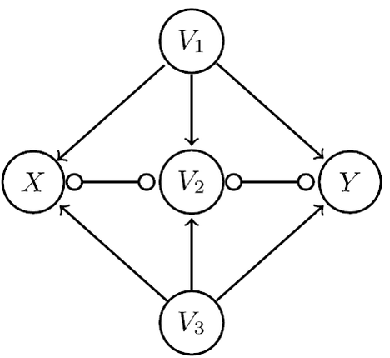}
 & \includegraphics{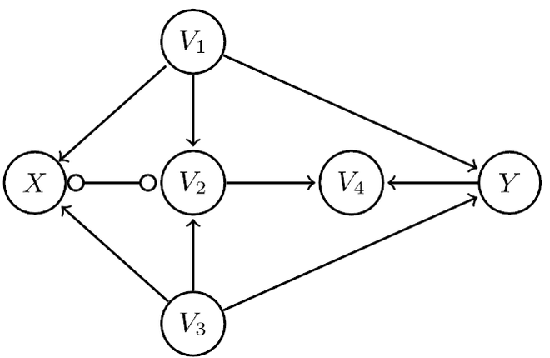}\\
\footnotesize{(a)} & \footnotesize{(b)}
\end{tabular}
\caption{CPDAG examples. \textup{(a)} The CPDAG $\mathcal C$ for Example~\protect\ref{exCPDAG1}.
\textup{(b)} The CPDAG $\mathcal{C}'$ for Example~\protect\ref{exCPDAG2}.}\label{figcpdag1a}\label{figcpdag1b}
\end{figure}

We now illustrate the theory for CPDAGs. In Example~\ref{exCPDAG2},
there is a set that satisfies the generalized back-door criterion,
while in Example~\ref{exCPDAG1} there is none.

\begin{example}\label{exCPDAG1}
    In the CPDAG $\mathcal C$ in Figure~\ref{figcpdag1a}(a), $f(y|\operatorname{do}(x))$ is
    not identifiable. To see this, note that the Markov equivalence class
    represented by this CPDAG contains three DAGs. Without loss of
    generality, we denote these by $\mathcal D_1, \mathcal D_2$ and~$\mathcal D_3$, where we assume that $\mathcal D_1$ contains the
    sub-graph $X \leftarrow V_2 \rightarrow Y$, $\mathcal D_2$ contains the
    sub-graph $X \leftarrow V_2 \leftarrow Y$, and $\mathcal{D}_3$ contains
    the sub-graph $X \rightarrow V_2 \rightarrow Y$. In $\mathcal D_1$ and
    $\mathcal D_2$ there is no directed path from $X$ to $Y$, so that
    $f(y|\operatorname{do}(x)) = f(y)$ for any density $f$ compatible with $\mathcal D_1$ or $\mathcal D_2$.
    In $\mathcal D_3$, however, there is a directed
    path from $X$ to $Y$. Hence, one can easily construct a density $f$ that is compatible with $\mathcal D_3$ such that $f(y|\operatorname{do}(x)) \neq f(y)$.
    This implies that $f(y|\operatorname{do}(x))$ is not identifiable. This implies that there cannot be a generalized back-door set relative to $(X,Y)$ and~$\mathcal C$.

    We now apply Theorem~\ref{thbackdoorsetforgeneralgraph} to the
    CPDAG $\mathcal C$ to check if this leads to the same
conclusion. Note that $\mathcal G = \mathcal C$ and $\mathcal R^* =
\{\mathcal D_3\}$. Hence, we take $\mathcal R = \mathcal D_3$ and the
corresponding $\mathcal R_{\underline X} = \mathcal D_3$. We then have
$\operatorname{D\mbox{-}SEP}(X,Y,\mathcal
    R_{\underline X}) = \{V_1,V_2,V_3\}$ and $\operatorname{possibleDe}(X,\mathcal
    G) = \{V_2,Y\}$. Hence, $\operatorname{D\mbox{-}SEP}
    (X,Y,\mathcal R_{\underline X}) \cap\break 
    \operatorname{possibleDe}(X,\mathcal G) =\{V_2\}$, and Theorem~\ref{thbackdoorsetforgeneralgraph} correctly says that it is impossible to satisfy the generalized back-door criterion
      relative to $(X,Y)$ and $\mathcal C$.

    Finally, we check if Corollary~\ref{corbackdoorsetCPDAG} also yields the same result. Note that $\mathcal C_{\underline X} = \mathcal C$ and $Y \in
    \operatorname{possibleDe}(X,\mathcal C_{\underline X}) = \{V_2,Y\}$. Hence,  we
    again find that it is impossible to satisfy the generalized back-door criterion relative to $(X,Y)$ and $\mathcal C$.
\end{example}

\begin{example}\label{exCPDAG2}
    In the CPDAG $\mathcal C'$ in Figure~\ref{figcpdag1b}(b), $f(y|\operatorname{do}(x))$
    is identifiable and equals $f(y)$, since there is no possibly directed
    path from $X$ to $Y$ in $\mathcal C'$.

    We now check if we also arrive at this conclusion by applying Theorem~\ref{thbackdoorsetforgeneralgraph}. Note that there are two DAGs
    in the Markov equivalence class described by~$\mathcal C'$, namely $\mathcal D'_1$ with the edge
    $X\to V_2$ and $\mathcal D'_2$ with the edge $X \leftarrow
V_2$. Thus in Theorem~\ref{thbackdoorsetforgeneralgraph}, we have
$\mathcal G = \mathcal C'$ and $\mathcal R^* = \{\mathcal D'_1\}$.
Hence we take $\mathcal R = \mathcal D'_1$ and the corresponding
$\mathcal R_{\underline X} = \mathcal D'_1$. Note that $Y\notin
    \operatorname{adj}(X,\mathcal R_{\underline X}) = \{V_1,V_2,V_3\}$ and
    $\operatorname{D\mbox{-}SEP}(X,Y,\mathcal R_{\underline{X}}) = \{V_1,V_3\}$ and
    $\operatorname{possibleDe}(X,\mathcal G) = \{V_2,V_4\}$.
    Hence,
    $\operatorname{D\mbox{-}SEP}(X,Y,\mathcal R_{\underline{X}}) \cap
    \operatorname{possibleDe}(X,\mathcal G) = \varnothing$, and
    $\operatorname{D\mbox{-}SEP}(X,Y,\mathcal R_{\underline{X}}) = \{V_1,V_3\}$\vspace*{1pt} satisfies
    the generalized back-door criterion relative to $(X,Y)$ and $\mathcal
    C'$. We can indeed check that the set $\{V_1,V_3\}$ satisfies the conditions in
    Definition~\ref{defbackdoorgraphical}.

    Finally, we also apply Corollary~\ref{corbackdoorsetCPDAG}. Note that $\mathcal C'_{\underline X} = \mathcal C'$. Moreover, $Y\notin
    \operatorname{pa}(X,\mathcal C')$ and $Y\notin
    \operatorname{possibleDe}(X,\mathcal C'_{\underline X})$. Hence, $\operatorname{pa}(X,\mathcal C') = \{V_1,V_3\}$ satisfies the generalized
    back-door criterion relative to $(X,Y)$ and $\mathcal C'$.
\end{example}

\subsection{MAG examples}

Next, we illustrate the theory for MAGs. In Examples \ref{exMAG1} and
\ref{exMAG2}, there does not exist a generalized back-door set relative
to $(X,Y)$ and the given MAGs. In Example~\ref{exMAG1}, this is due to
$Y\in \operatorname{adj}(X, \mathcal M_{\underline X})$, while in Example~\ref{exMAG2}, it is due to $\operatorname{D\mbox{-}SEP}(X,Y,\mathcal M_{\underline X})
\cap \operatorname{de}(X,\mathcal M) \neq \varnothing$.

\begin{figure}
\begin{tabular}{@{}cc@{}}

\includegraphics{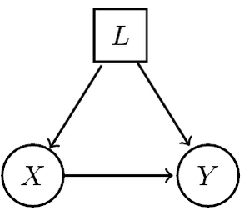}
 & \includegraphics{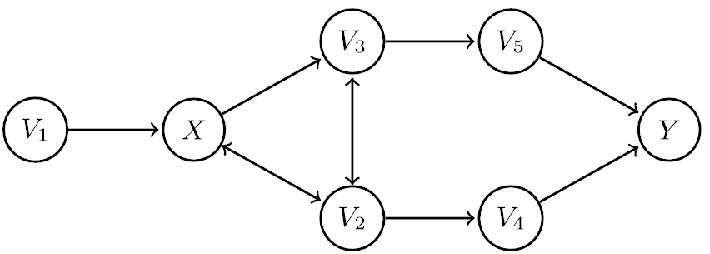}\\
\footnotesize{(a)} & \footnotesize{(b)}
\end{tabular}
\caption{MAG examples. \textup{(a)} A possible DAG described by the MAG in Example~\protect\ref{exMAG1},
    where $L$ is latent. \textup{(b)} The MAG $\mathcal M$ for Example~\protect\ref{exMAG2}.}\label{figMAG2}\label{figMAG1->DAG}
\end{figure}

\begin{example}\label{exMAG1}
   Consider the MAG $\mathcal M$ consisting of the invisible edge $X
   \rightarrow Y$, and suppose we are interested in $f(y|\operatorname{do}(x))$. Then underlying DAG could be as in Figure~\ref{figMAG1->DAG}(a), where $L$ is unobserved. This is a well-known
   example where $f(y|\operatorname{do}(x))$ is not identifiable.

   We now apply Corollary~\ref{corbackdoorsetMAG} to check if we indeed
   find that it is impossible to satisfy the generalized back-door criterion relative to $(X,Y)$ and $\mathcal M$.
   We have that $\mathcal M = \mathcal M_{\underline X}$ is the graph $X \to Y$.
   Hence, $Y\in \operatorname{adj}(X,\mathcal M_{\underline
     X})$, which leads to the correct conclusion.
\end{example}

\begin{example}\label{exMAG2}
  Consider the MAG $\mathcal M$ in Figure~\ref{figMAG2}(b) and apply Corollary~\ref{corbackdoorsetMAG} with $\mathbf{X}=\{X\}$ and $\mathbf{Y} = \{Y\}$. Since the edge
  $X\to V_3$ is visible, $\mathcal M_{\underline X}$ is constructed from
  $\mathcal M$ by removing this edge. We then have
  $\operatorname{D\mbox{-}SEP}(X,Y,\break \mathcal M_{\underline X}) = \{V_1,V_2,V_3\}$ and
  $\operatorname{de}(X,\mathcal M) = \{V_3,V_5,Y\}$. Hence the intersection of
  $\operatorname{de}(X,\mathcal M)$ and $\operatorname{D\mbox{-}SEP}(X,Y,\mathcal M_{\underline X})$ is nonempty, and it follows that there is no generalized back-door set
  relative to $(X,Y)$ and $\mathcal M$.

  Indeed, we see that it is impossible to satisfy conditions (B-i) and
  (B-ii) in Definition~\ref{defbackdoorgraphical}. In order to block the
  back-door path $\langle X, V_2, V_4,Y\rangle$, we must include $V_2$
  or $V_4$ in our set $\mathbf{W}$, but doing so opens the collider $V_2$ on the
  back-door path $\langle X, V_2, V_3, V_5, Y\rangle$. Hence, the latter path
  must be blocked by $V_3$ or $V_5$. But both these vertices are
  descendants of $X$ in $\mathcal M$, and are therefore not allowed by condition
  (B-i).
\end{example}

\subsection{PAG example}

Finally, Example~\ref{exPAG} is an example where there exists a
generalized back-door set relative to some $(X,Y)$ and a PAG. This
example also illustrates that there may be subsets of
$\operatorname{D\mbox{-}SEP}(X,Y,\mathcal R_{\underline X})$ in Theorem~\ref{thbackdoorsetforgeneralgraph} that satisfy the generalized
back-door criterion. In other words, Theorem~\ref{thbackdoorsetforgeneralgraph} may yield a nonminimal set. Hence,
if one is interested in a minimal generalized back-door set, one could
consider all subsets of $\operatorname{D\mbox{-}SEP}(X,Y,\mathcal R_{\underline X})$.
Example~\ref{exPAG} also illustrates why $\mathcal R_{\underline X}$ is
required to satisfy Definition~\ref{defR}.

\begin{figure}
\begin{tabular}{@{}cc@{}}

\includegraphics{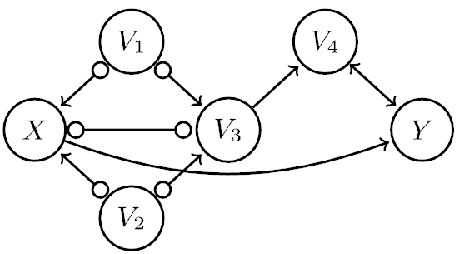}
 & \includegraphics{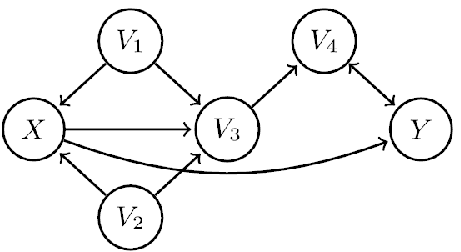}\\
\footnotesize{(a)} & \footnotesize{(b)}
\end{tabular}
\caption{PAG example. \textup{(a)} The PAG $\mathcal{P}$ for Example~\protect\ref{exPAG}. \textup{(b)} A possible MAG $\mathcal{M}$ for
Example~\protect\ref{exPAG}.}\label{figPAG}\label{figPAG->MAG}
\end{figure}

\begin{example}\label{exPAG}
  Consider the PAG $\mathcal P$ in Figure~\ref{figPAG}(a), and suppose we are interested in $f(y|\operatorname{do}(x))$.
  Note that the MAG $\mathcal R = \mathcal M$ as given in Figure~\ref{figPAG->MAG}(b) is in~$\mathcal R^*$; see Definition~\ref{defR}.
  We will apply Theorem~\ref{thbackdoorsetforgeneralgraph} using the corresponding graph $\mathcal R_{\underline X}$, which
   is as $\mathcal M$ but without the edge $X \to Y$.
  We then have $Y\notin \operatorname{adj}(X,\mathcal R_{\underline X})$ and
   $\operatorname{D\mbox{-}SEP}(X,Y,\mathcal R_{\underline X}) \cap
   \operatorname{possibleDe}(X,\mathcal G) = \{V_1,V_2\} \cap \{V_3,V_4,Y\} =
   \varnothing$. Hence Theorem~\ref{thbackdoorsetforgeneralgraph}
   implies that $\{V_1,V_2\}$ is a generalized back-door
   set relative to $(X,Y)$ and~$\mathcal P$. One can easily verify
   that all subsets of $\{V_1,V_2\}$ are also generalized back-door
   sets relative to $(X,Y)$ and $\mathcal P$, since all back-door paths from $X$ to $Y$
   are blocked by the collider $V_4$ on these paths. This shows that
   $\operatorname{D\mbox{-}SEP}(X,Y,\mathcal R_{\underline X})$ is not minimal.

   This example also shows the importance of Definition~\ref{defR}. To see this, let $\mathcal
   R'$ be as $\mathcal R$, but with the edge $X\leftarrow V_3$ instead of $X\to
   V_3$, so that there is an additional edge into $X$. Then
   $\operatorname{D\mbox{-}SEP}(X,Y,\mathcal R'_{\underline X}) = \{V_1,V_2,V_3\}$, and we
   get $\operatorname{D\mbox{-}SEP}(X,Y,\mathcal R'_{\underline X}) \cap
   \operatorname{possibleDe}(X,\mathcal G) = \{V_3\} \neq \varnothing$.\vspace*{1pt} This shows
   that applying Theorem~\ref{thbackdoorsetforgeneralgraph} with
   $\mathcal R'_{\underline X}$ instead of $\mathcal R_{\underline X}$ leads to incorrect results.
\end{example}

\section{Discussion}\label{secdiscussion}

In this paper, we generalize Pearl's back-door criterion
[\citet{Pearl93}] to a generalized back-door criterion for DAGs,
CPDAGs, MAGs and PAGs. We also provide easily checkable necessary and
sufficient criteria for the existence of a generalized back-door set,
when considering a single intervention variable and a single outcome
variable. Moreover, if such a set exists, we provide an explicit set
that satisfies the generalized back-door criterion. This set is not
necessarily minimal, so if one is interested in a minimal set, one
could consider all subsets.

Although effects that can be computed via the generalized back-door
criterion are only a subset of all identifiable causal effects, we hope
that the generalized back-door criterion will be useful in practice,
and will make it easier to work with CPDAGs, MAGs and PAGs. Moreover,
combining our results for CPDAGs and PAGs with fast causal structure
learning algorithms such as the PC algorithm [\citet{SpirtesEtAl00}] or
the FCI algorithm
[\citet{SpirtesEtAl00,ColomboEtAl12,ClaassenMooijHeskes13}] yields a
computationally efficient way to obtain information on causal effects
when assuming that the observational distribution is faithful to the
true unknown causal DAG with or without hidden variables. To our
knowledge, the prediction algorithm of \citet{SpirtesEtAl00} is the only
alternative approach under the same assumptions, but the prediction
algorithm is computationally much more complex.

The IDA algorithm
[\citet{MaathuisKalischBuehlmann09,MaathuisColomboKalischBuehlmann10}]
has been designed to obtain bounds on causal effects when assuming that
the observational distribution is faithful to the true underlying
causal DAG without hidden variables. IDA roughly combines the PC
algorithm with Pearl's back-door criterion. We could now apply a
similar approach in the setting with hidden variables, by combining the
FCI algorithm with the generalized back-door criterion for MAGs.

Possible directions for future work include studying the exact relationship between the prediction algorithm and our generalized back-door criterion, 
generalizing the results in Section~\ref{secfindingbackdoorset} to
allow for sets $\mathbf{X}$ and $\mathbf{Y}$ and extending the recent
results of \citet{VanderZanderEtAl14} to CPDAGs and PAGs.

\section{Proofs}\label{secproofs}

\subsection{Proofs for Section 
\texorpdfstring{\protect\ref{secgeneralizedbackdoorcriterion}}{3}}

In order to prove Theorem~\ref{thgeneralizedbackdoorsufficientforadjustment}, we formulate
so-called invariance conditions that will turn out to be sufficient for
adjustment; see Definition~\ref{definvariancecriterion} and Theorem~\ref{thinvarianceimpliesadjustment} below. First, we briefly define
what is meant by invariance. We refer to
\citet{Zhang08-causal-reasoning-ancestral-graphs} for full details.

Let $\mathbf{Y}$, $\mathbf{Z}$ and $\mathbf{X}$ be three subsets of
vertices in a causal DAG $\mathcal D$, where $\mathbf{X}\cap \mathbf{Y}
= \mathbf{Y}\cap \mathbf{Z} = \varnothing$. Then a density
$f(\mathbf{y}|\mathbf{z})$ is said to be entailed to be invariant under
interventions on $\mathbf{X}$ given $\mathcal D$ if
$f_{\mathbf{X}:=\mathbf{x}}(\mathbf{y}|\mathbf{z}) =
f(\mathbf{y}|\mathbf{z})$ for all causal Bayesian networks $(\mathcal
D, f)$, where the subscript $\mathbf{X}:=\mathbf{x}$ denotes
$\operatorname{do}(\mathbf{X}=\mathbf{x})$. (This notation is used since $\mathbf{X}$
and $\mathbf{Z}$ are allowed to overlap.) The density
$f(\mathbf{y}|\mathbf{z})$ is said to be entailed to be invariant under
interventions on $\mathbf{X}$ given a CPDAG $\mathcal C$, a MAG
$\mathcal M$  or a PAG $\mathcal P$ if it is entailed to be invariant
under interventions on $\mathbf{X}$ given all DAGs represented by
$\mathcal C$,  $\mathcal M$ or $\mathcal P$, respectively.

\begin{definition}[(Invariance criterion)]\label{definvariancecriterion}
   Let $\mathbf{X}$, $\mathbf{Y}$ and $\mathbf{W}$ be pairwise disjoint sets of
   vertices in $\mathcal{G}$, where $\mathcal{G}$ is a DAG, CPDAG, MAG or PAG. Then $\mathbf{W}$ satisfies the \textit{invariance criterion} relative to
   $(\mathbf{X},\mathbf{Y})$ and $\mathcal G$ if the following two conditions hold for any density $f$ compatible with $\mathcal G$:
  \begin{longlist}[(I-ii)]
     \item[{(I-i)}]
      $f(\mathbf{w}|\operatorname{do}(\mathbf{x})) = f(\mathbf{w})$;
     \item[{(I-ii)}] $f(\mathbf{y}|\operatorname{do}(\mathbf{x}),\mathbf{w}) = f(\mathbf{y}|\mathbf{x},\mathbf{w})$.
  \end{longlist}
\end{definition}

In other words, conditions (I-i) and (I-ii) state that $f(\mathbf{w})$
and $f(\mathbf{y}|\mathbf{x},\mathbf{w})$ are entailed to be invariant
under interventions on $\mathbf{X}$ given $\mathcal G$. The conditions
are also closely related to the conditions in equation (9) of
\citet{Pearl93}. We note that condition (I-i) is trivially satisfied if
$\mathbf{W}=\varnothing$.

\begin{theorem}\label{thinvarianceimpliesadjustment}
  Let $\mathbf{X}$, $\mathbf{Y}$ and $\mathbf{W}$ be pairwise disjoint sets of vertices
  in $\mathcal G$, where $\mathcal G$ is a DAG, CPDAG, MAG or PAG. If
  $\mathbf{W}$ satisfies the invariance criterion relative to
  $(\mathbf{X},\mathbf{Y})$ and $\mathcal G$,
  then it satisfies the adjustment criterion relative to
  $(\mathbf{X},\mathbf{Y})$ and $\mathcal G$.
\end{theorem}

\begin{pf}
  If $\mathbf{W}=\varnothing$, condition (I-ii) immediately gives 
  $f(\mathbf{y}|\operatorname{do}(\mathbf{x})) = \break f(\mathbf{y}|\mathbf{x})$. Otherwise, we have
%
\begin{equation}
\label{eq1}
\hspace*{4pt} f\bigl(\mathbf{y}|\operatorname{do}(\mathbf{x})\bigr) = \int_{\mathbf{w}}
f\bigl(\mathbf{y},\mathbf{w}|\operatorname{do}(\mathbf{x})\bigr)\,d\mathbf{w} = \int
_{\mathbf{w}} f\bigl(\mathbf{y}|\mathbf{w},\operatorname{do}(\mathbf{x})\bigr)f\bigl(
\mathbf{w}|\operatorname{do}(\mathbf{x})\bigr)\,d\mathbf{w}.
\end{equation}
   Under conditions (I-i) and (I-ii), the right-hand side of \eqref{eq1}
   simplifies to $\int_{\mathbf{w}} f(\mathbf{y}|\mathbf{w}, \mathbf{x})
   f(\mathbf{w})\,d\mathbf{w}$.
\end{pf}

\citeauthor{SpirtesEtAl93} (\citeyear{SpirtesEtAl93,SpirtesEtAl00}), \citet{Zhang08-causal-reasoning-ancestral-graphs}
formulated invariance results for DAGs, MAGs and PAGs. We derive a
similar result for CPDAGs and then summarize the results for all these
types of graphs in Theorem~\ref{thinvariancegeneral}.
%

\begin{theorem}[(Graphical criteria for invariance)]\label{thinvariancegeneral}
Let $\mathbf{X},\mathbf{Y},\mathbf{Z}$ be three subsets of observed
vertices in $\mathcal G$, where $\mathcal G$ represents a DAG, CPDAG,
MAG or PAG. Moreover, let  $\mathbf{X} \cap \mathbf{Y} = \mathbf{Y}
\cap \mathbf{Z} = \varnothing$. Then $f(\mathbf{y}|\mathbf{z})$ is
entailed to be invariant under interventions on $\mathbf{X}$ given
$\mathcal G$  if and only if:
   \begin{longlist}[(3)]
   \item[(1)] for every $X \in \mathbf{X}\cap \mathbf{Z}$, every
       m-connecting definite status path, if any, between $X$ and
       any member of $\mathbf{Y}$ given $\mathbf{Z}\setminus \{X\}$
       is out of $X$ with a visible edge;
       \item[(2)] for every $X \in \mathbf{X} \cap
           (\operatorname{possibleAn}(\mathbf{Z}, \mathcal G)\setminus
           \mathbf{Z})$, there is no m-connecting definite status
           path between $X$ and any member of $\mathbf{Y}$ given
           $\mathbf{Z}$;
   \item[(3)] for every $X \in \mathbf{X}\setminus
       \operatorname{possibleAn}(\mathbf{Z}, \mathcal G)$, every
       m-connecting definite status path, if any, between $X$ and
       any member of $\mathbf{Y}$ given $\mathbf{Z}$ is into $X$.
   \end{longlist}
\end{theorem}

\begin{pf}
One can easily check that the conditions reduce to the appropriate
conditions for DAGs, MAGs and PAGs
[\citet{Zhang08-causal-reasoning-ancestral-graphs}, Proposition~18, Theorem~24 and Theorem~30]. The result for CPDAGs can be proved
analogously.
\end{pf}

Note that $\mathbf{X}\cap \mathbf{Z}$, $\mathbf{X}\cap
(\operatorname{possibleAn}(\mathbf{Z}, \mathcal G)\setminus \mathbf{Z})$ and
$\mathbf{X}\setminus \operatorname{possibleAn}(\mathbf{Z}, \mathcal G)$ form a
partition of $\mathbf{X}$. Hence, only one of the conditions in Theorem~\ref{thinvariancegeneral} is relevant for a given $X \in \mathbf{X}$.

%
%

We also need the following basic property of PAGs and CPDAGs:

\begin{lemma}[{[Basic property of CPDAGs and PAGs;
Lemma~1 of \citet{Meek95} for CPDAGs, and Lemma~3.3.1 of
\citet{Zhang06-dissertation} for
PAGs]}]\label{lemmapropertyCP1forCPDAGsandPAGs}
%
%
For any three vertices $A$, $B$ and $C$ in a CPDAG $\mathcal C$ or PAG
$\mathcal P$, the following holds: if $A \stararrow B \circstar C$,
then there is an edge between $A$ and $C$ with an arrowhead at $C$,
namely $A \stararrow C$. Furthermore, if the edge between $A$ and $B$
is $A \to B$, then the edge between $A$ and $C$ is either $A \circarrow
C$ or $A \rightarrow C$ (i.e., not $A \leftrightarrow C$).
\end{lemma}

We now show that the invariance conditions in Definition~\ref{definvariancecriterion} are equivalent to the graphical conditions
of Definition~\ref{defbackdoorgraphical}.

\begin{theorem}\label{thbackdoorequivalenttoinvariance}
   The generalized back-door criterion (Definition~\ref{defbackdoorgraphical}) is equivalent to the invariance criterion (Definition~\ref{definvariancecriterion}).
\end{theorem}

\begin{pf}
We first show that condition (B-ii) of Definition~\ref{defbackdoorgraphical} is equivalent to condition (I-ii) of
Definition~\ref{definvariancecriterion}. We use Theorem~\ref{thinvariancegeneral} with $(\mathbf{X'},
\mathbf{Y'},\mathbf{Z'})$, where $\mathbf{X'}=\mathbf{X}$,
$\mathbf{Y'}=\mathbf{Y}$ and $\mathbf{Z'} = \mathbf{X}\cup\mathbf{W}$.
Then $\mathbf{X'} \subseteq \mathbf{Z'}$, and clause (1) of the theorem
yields that (I-ii) is equivalent to the following: for every $X\in
\mathbf{X}$, every m-connecting definite status path, if any, between
$X$ and any member of $\mathbf{Y}$ given $(\mathbf{X}
  \cup \mathbf{W}) \setminus\{X\}$ is out of $X$ with a visible edge.
  This is equivalent to condition~(B-ii) by our definition of a back-door path; see Definition~\ref{defbackdoorpath}.

By Lemma~\ref{lemmapossibledescendantsdefinitestatuspath} (below),
condition (B-i) of Definition~\ref{defbackdoorgraphical} is equivalent
to condition (B-i)$'$ of Remark~\ref{rempossibledescendantsdefinitestatuspath}.

We now show that condition (B-i)$'$ is equivalent to condition (I-i) in
Definition~\ref{definvariancecriterion}. We use Theorem~\ref{thinvariancegeneral} with $(\mathbf{X'},
\mathbf{Y'},\mathbf{Z'})$, where $\mathbf{X'}=\mathbf{X}$,
$\mathbf{Y'}=\mathbf{W}$ and $\mathbf{Z'}=\varnothing$. Then
$\mathbf{Z'}=\operatorname{possibleAn}(\mathbf{Z'},\mathcal G)=\varnothing$ and
clause (3) of the theorem yields that (I-i) is equivalent to the
following condition (I-i)$'$:
  for every $X \in \mathbf{X}$, every m-connecting definite status path, if any, between $X$ and any member
  of $\mathbf{W}$ is into $X$. We now show that (I-i)$'$ is equivalent to (B-i)$'$.

  First suppose
   that $\mathbf{W}$ violates (B-i)$'$. Then there are $W\in
   \mathbf{W}$ and $X \in \mathbf{X}$ such that there is a possibly directed definite status path $p$
   from $X$ to $W$. Since $p$ is possibly directed, it is not into $X$ and it cannot contain colliders.
   Hence, it is an m-connecting definite status path between $X$ and $W$ that is not into $X$. This violates (I-i)$'$.

Now suppose that $\mathbf{W}$ violates (I-i)$'$. Then there are $W \in \mathbf{W}$ and $X\in \mathbf{X}$ such that there
is an m-connecting definite status path between $X$ and $W$ that is not into $X$. Let $p = \langle X=U_1, \dots, U_k=W\rangle$ be such a path.
Then every nonendpoint vertex on $p$ must be a definite noncollider. Suppose that $p$ is not a possibly directed path from $X$ to $W$,
meaning that there exists an $i \in \{2,\dots,k\}$ such that the edge between $U_{i-1}$ and $U_i$ is into $U_{i-1}$. If $i=2$, this
means that the path is into $X$, which is a contradiction. If $i>2$, then the edge between $U_{i-2}$ and $U_{i-1}$ must be out of $U_{i-1}$, since $U_{i-1}$ is a definite
noncollider. But this means that the edge must be into $U_{i-2}$, since edges of the form $\circtail$ or $\tailtail$ are not allowed. Continuing this argument,
we find that for all $j\in \{2,\dots, i\}$, the edge between $U_{j-1}$ and $U_j$ is into $U_{j-1}$. But this means that the path is into $U_1 = X$, which is a
contradiction. Hence, $p$ is a possibly directed path from $X$ to $W$. Together with the fact that $p$ is of a definite status, this violates (B-i)$'$.
\end{pf}

\begin{lemma}\label{lemmapossibledescendantsdefinitestatuspath}
   Let $X$ and $Y$ be two distinct vertices in $\mathcal G$, where $\mathcal G$ is a DAG, CPDAG, MAG or PAG.
   If $Y \in \operatorname{possibleDe}(X,\mathcal G)$, then there is a possibly directed definite status path $p= \langle X = U_1,\ldots,U_k=Y \rangle$ from $X$ to $Y$.
   Moreover, if $U_{i-1} \stararrow U_i$ for some $i\in \{2,\ldots,k\}$, then $U_{j-1} \to U_{j}$ for all $j\in \{i+1,\ldots,k\}$.
\end{lemma}

\begin{pf}
   If $\mathcal G$ is a DAG or a MAG, the lemma is trivially true. So let $\mathcal G$ be a CPDAG or a PAG, and assume that
   $Y\in \operatorname{possibleDe}(X,\mathcal G)$. This implies that there is a possibly directed path from $X$ to $Y$ in
   $\mathcal G$. Let $p = \langle X=U_1,\dots,U_k=Y\rangle$ be a shortest such path. If $p$ is of length one, then the Lemma is trivially true.
   So assume that the length of $p$ is at least two, that is, $k\ge 3$.

We first show that $p$ is a definite status path. Note that $p$ can
contain the following edges $U_{i-1} \circcirc U_i$, $U_{i-1}
\circarrow U_{i}$ and $U_{i-1} \rightarrow U_{i}$ ($i=2,\ldots,k$).  We
now consider a sub-path $p(U_{i-1},U_{i+1}) = \langle
U_{i-1},U_i,U_{i+1} \rangle$ of $p$, for some $i\in\{2,\ldots,k-1\}$.

This sub-path cannot be of the form $U_{i-1} \circarrow\break  U_i \circcirc
U_{i+1}$ or $U_{i-1} \circarrow U_i \circarrow U_{i+1}$. To see this,
suppose that the sub-path takes such a form. Then Lemma~\ref{lemmapropertyCP1forCPDAGsandPAGs} implies the edge $U_{i-1}
\stararrow U_{i+1}$. Suppose that this edge is into $U_{i-1}$; that is,
it is $U_{i-1} \leftrightarrow U_{i+1}$. Then Lemma~\ref{lemmapropertyCP1forCPDAGsandPAGs} applied to $U_{i+1}
\leftrightarrow U_{i-1} \circarrow U_i$ implies the edge $U_{i+1}
\stararrow U_i$, which is a contradiction. If the edge $U_{i-1}
\stararrow U_{i+1}$ is not into $U_{i-1}$, then $p$ is not a shortest
possibly directed path.

Similarly, the sub-path cannot be of the form $U_{i-1} \to U_i
\circcirc U_{i+1}$ or $U_{i-1} \rightarrow U_i \circarrow U_{i+1}$. To
see this, suppose that the sub-path takes such a form. Then Lemma~\ref{lemmapropertyCP1forCPDAGsandPAGs} implies the edge $U_{i-1}
\circarrow U_{i+1}$ or $U_{i-1} \to U_{i+1}$. In either case, $p$ is
not a shortest possibly directed path.

Moreover, if the sub-path is of the form $U_{i-1} \circcirc U_i
\circcirc U_{i+1}$, $U_{i-1} \circcirc U_i \circarrow\!\break  U_{i+1}$ or
$U_{i-1} \circcirc U_i \to U_{i+1}$, then it must be unshielded. To see
this, suppose that the sub-path takes such a form and is not
unshielded. If the edge between $U_{i-1}$ and $U_{i+1}$ is into
$U_{i-1}$, then Lemma~\ref{lemmapropertyCP1forCPDAGsandPAGs} applied to
$U_{i+1} \stararrow U_{i-1} \circcirc\break  U_i$ implies the edge $U_{i+1}
\stararrow U_i$, which is a contradiction. If the edge between
$U_{i-1}$ and $U_{i+1}$ is not into $U_{i-1}$, then $p$ is not a
shortest possibly directed path.

Hence, $p$ can only contain triples of the form $U_{i-1} \circarrow U_i
\to U_{i+1}$ or $U_{i-1} \to U_i \to U_{i+1}$, or of the form $U_{i-1}
\circcirc U_i \circcirc U_{i+1}$, $U_{i-1} \circcirc U_i \circarrow
U_{i+1}$ or $U_{i-1} \circcirc\!\break   U_i \to U_{i+1}$ where $U_{i-1}$ and
$U_{i+1}$ are not adjacent. In all these cases, the middle vertex $U_i$
is a definite noncollider, so that $p$ is a definite status path.
Finally, if $U_{i-1} \stararrow U_i$ for some $i\in \{2,\dots,k\}$, it
follows that $U_{j-1} \to U_{j}$ for all $j\in \{i+1,\dots,k\}$.
\end{pf}

\begin{pf*}{Proof of Theorem \protect\ref{thgeneralizedbackdoorsufficientforadjustment}}
   This follows directly from Theorems \ref{thinvarianceimpliesadjustment} and \ref{thbackdoorequivalenttoinvariance}.
\end{pf*}

\begin{pf*}{Proof of Lemma \protect\ref{lemmaourbackdoorstrongerthanPearls}}
Conditions (P-i) and (B-i) are trivially equivalent for DAGs. We
therefore only show that (P-ii) implies (B-ii), by contradiction. Thus,
suppose that $\mathbf{W}$ blocks all back-door paths between $X\in
\mathbf{X}$ and $Y\in \mathbf{Y}$ in $\mathcal{D}$, but there exist $X \in
\mathbf{X}$ and $Y \in \mathbf{Y}$ such that there is a back-door path
$p$ from $X$ to $Y$ that is not blocked by
$\mathbf{W}\cup\mathbf{X}\setminus\{X\}$. This means that: (i) no
noncollider on $p$ is in $\mathbf{W}\cup\mathbf{X}\setminus \{X\}$,
(ii) all colliders on $p$ have a descendant in $\mathbf{W}\cup
\mathbf{X}\setminus\{X\}$, (iii) there is at least one collider on $p$
that has a descendant in $\mathbf{X} \setminus \{X\}$ but not in
$\mathbf{W}$. Among all colliders satisfying (iii), let $Q$ be the one
that is closest to $Y$ on $p$, and let $X'$ denote a descendant of $Q$
in $\mathbf{X}\setminus\{X\}$. Then the directed path $q(Q,X')$ from
$Q$ to $X'$ is m-connecting given $\mathbf{W}$, since it is a path
consisting of noncolliders and none of its vertices are in
$\mathbf{W}$.  Moreover, the sub-path $p(Q,Y)$ of $p$ is m-connecting
given $\mathbf{W}$ by construction. But this means that $q(X',Q) \oplus
p(Q,Y)$ is a back-door path from $X'$ to $Y$ that is m-connecting given
$\mathbf{W}$. This contradicts (P-ii).
\end{pf*}

\subsection{Proofs for Section 
\texorpdfstring{\protect\ref{secfindingbackdoorset}}{4}}

We first give several lemmas, starting with a result about m-connection
in MAGs. This result basically says that replacing condition (b) in
Definition~\ref{defm-connection} by ``every collider on the path is an
ancestor of some member of $\mathbf{Z} \cup\{X,Y\}$'' does not change
the m-separation relations in a MAG.

\begin{lemma}[{[\citet{Richardson03}, Corollary~1]}]
\label{lemmaRichardsoncorollary}
Let $X$ and $Y$ be two distinct vertices and $\mathbf{Z}$ be a subset
of vertices in a mixed graph $\mathcal M$, with $ \mathbf{Z}\cap
\{X,Y\} = \varnothing$. If there is a path between $X$ and $Y$ in
$\mathcal M$ on which no noncollider is in $\mathbf{Z}$ and every
collider is in $\operatorname{an}(\mathbf{Z} \cup \{X,Y\},\mathcal M)$, then
there is a path (not necessarily the same path) m-connecting $X$ and
$Y$ given $\mathbf{Z}$ in $\mathcal M$.
\end{lemma}

\begin{pf*}{Proof of  Lemma \protect\ref{lemmaD-SEP}}
Let $\mathcal G$ be an ancestral graph. First, we note that (iii)
trivially implies (i). Next, we show that (i) implies (ii), or
equivalently, that not (ii) implies not (i). Thus, suppose that $Y\in
\operatorname{D\mbox{-}SEP}(X,Y,\mathcal G)$. Then there is a collider path between
$X$ and $Y$ such that every vertex on the path is an ancestor of
$\{X,Y\}$ in $\mathcal G$. This path is m-connecting given any subset
of the remaining vertices, by Lemma~\ref{lemmaRichardsoncorollary}.

Next, we show that (ii) implies (iii). Suppose that $Y\notin
\operatorname{D\mbox{-}SEP}(X,Y,\mathcal G)$. If there is no path between $X$ and $Y$
in $\mathcal G$, then $X$ and $Y$ are trivially m-separated by any
subset of the remaining vertices. Thus, assume that there is at least
one path between $X$ and $Y$. Consider an arbitrary such path, and call
it $p$. Since $Y\notin \operatorname{D\mbox{-}SEP}(X,Y,\mathcal G)$, we have $Y\notin
\operatorname{adj}(X,\mathcal G)$. Hence the length of $p$ must be at least
two. We will show that $p$ is blocked by $\operatorname{D\mbox{-}SEP}(X,Y,\mathcal
G)$.

Suppose $p$ starts with $X \leftarrow V$. Then $V\in
\operatorname{D\mbox{-}SEP}(X,Y,\mathcal G)$, since $V \in \operatorname{an}(X,\mathcal G)$.
Since  $V$ is a noncollider on $p$, this implies that $p$ is blocked
by $\operatorname{D\mbox{-}SEP}(X,Y,\mathcal G)$.

Suppose $p$ is of the form $X \stararrow V \to \cdots \to Y$. Then $V\in
\operatorname{an}(Y,\mathcal G)$, so that $V\in \operatorname{D\mbox{-}SEP}(X,Y,\mathcal G)$.
Since $V$ is a noncollider on $p$, this implies that $p$ is blocked by
$\operatorname{D\mbox{-}SEP}(X,Y,\mathcal G)$.

Suppose $p$ starts with $X \stararrow V \to \cdots$ and the sub-path
$p(V,Y)$ of $p$ contains at least one collider. Let $C$ be the collider
closest to $V$ on $p$. Then $V \in \operatorname{an}(C,\mathcal G)$. If $C
\notin\operatorname{an}(\operatorname{D\mbox{-}SEP}(X,Y,\mathcal G), \mathcal G)$, then $p$ is
blocked by $\operatorname{D\mbox{-}SEP}(X,Y,\mathcal G)$. Hence, suppose $C \in
\operatorname{an}(\operatorname{D\mbox{-}SEP}(X,Y,\mathcal G), \mathcal G)$. 
Since any vertex
in $\operatorname{D\mbox{-}SEP}(X,Y,\break \mathcal G)$ is an ancestor of $\{X,Y\}$ in
$\mathcal G$, this implies $C \in \operatorname{an}(\{X,Y\},\mathcal G)$ and
hence $V \in \operatorname{an}(\{X,Y\},\mathcal G)$ and $V \in
\operatorname{D\mbox{-}SEP}(X,Y,\mathcal G)$. Since $V$ is a noncollider on $p$, $p$
is blocked by $\operatorname{D\mbox{-}SEP}(X,Y,\mathcal G)$.

Suppose $p$ is a collider path of the form $X \stararrow
\leftrightarrow \cdots \arrowstar Y$. Then at least one of the
colliders is not in $\operatorname{an}(\{X,Y\},\mathcal G)$, since otherwise
$Y\in \operatorname{D\mbox{-}SEP}(X,Y,\break \mathcal G)$. Let $C$ be the collider closest to
$X$ on $p$ that is not in $\operatorname{an}(\{X,Y\},\mathcal G)$. Then $C
\notin \operatorname{an}(\operatorname{D\mbox{-}SEP}(X,Y,\mathcal G),\mathcal G)$. Hence, $p$
is blocked by $\operatorname{D\mbox{-}SEP}(X,Y,\mathcal G)$.

Suppose $p$ is of the form $X \stararrow \leftrightarrow \cdots
\leftrightarrow V \leftarrow W \cdots Y$, with $W\neq Y$ ($W=Y$ was
treated in the previous case) and the sub-path $p(X,V)$ is allowed to
be of length one (i.e., $X \stararrow V$). If $W \in
\operatorname{D\mbox{-}SEP}(X,Y,\mathcal G)$, then $p$ is blocked by
$\operatorname{D\mbox{-}SEP}(X,Y,\mathcal G)$. So suppose that  $W\notin
\operatorname{D\mbox{-}SEP}(X,Y,\mathcal G)$. Then there does not exist a collider
path between $X$ and $W$ such that each vertex on the path is in
$\operatorname{an}(\{X,Y\},\mathcal G)$. This implies that there is a collider
on the sub-path $p(X,W)$ of $p$ that is not in
$\operatorname{an}(\{X,Y\},\mathcal G)$. Among such vertices, let $Z$ be the
one that is closest to $X$ on $p(X,W)$. Then $Z \notin
\operatorname{an}(\operatorname{D\mbox{-}SEP}(X,Y,\mathcal G),\mathcal G)$. Hence, $p$ is
blocked by $\operatorname{D\mbox{-}SEP}(X,Y,\mathcal G)$.

Finally, if $\mathcal G$ is a MAG, two vertices are adjacent if and
only if no subset of the remaining variables can m-separate them.
Hence, (i) and (iv) are equivalent for MAGs.
\end{pf*}

The following lemma says that we can check the existence of
m-connecting definite status back-door paths in $\mathcal{G}$ by checking
the existence of m-connecting paths in $\mathcal R_{\underline X}$,
where $\mathcal R_{\underline X}$ is any graph satisfying Definition~\ref{defR}. This lemma is closely related to Lemma~5.1.7 of
\citet{Zhang06-dissertation} and Lemmas 26 and 27 of
\citet{Zhang08-causal-reasoning-ancestral-graphs}.

\begin{lemma}\label{lemmam-sepPAG2MAG}
Let $X$ and $Y$ be two distinct vertices and $\mathbf{Z}$ be a subset
of vertices in $\mathcal G$, where $\mathcal G$ is a DAG, CPDAG, MAG or
PAG. Let $\mathcal R_{\underline X}$ be any graph satisfying Definition~\ref{defR}. Then there is a definite status m-connecting back-door path
from $X$ to $Y$ given $\mathbf{Z}$ in $\mathcal G$ if and only if there
is an m-connecting path between $X$ and $Y$ given $\mathbf{Z}$ in
$\mathcal R_{\underline X}$.
\end{lemma}

\begin{pf}
Let $\mathcal R \in \mathcal R^*$ and $\mathcal R_{\underline X}$
satisfy Definition~\ref{defR}. We first prove the ``only if'' statement.
Suppose there is a definite status m-connecting back-door path $p$ from
$X$ to $Y$ given $\mathbf{Z}$ in $\mathcal G$. Let $p'$ and $p''$ be
the corresponding paths in $\mathcal R$ and $\mathcal R_{\underline
X}$, consisting of the same sequence of vertices. (Note that $p''$
exists by the definition of $\mathcal R_{\underline X}$ and the fact
that $p$ is a back-door path in $\mathcal G$.) Then the path $p'$ is
m-connecting given $\mathbf{Z}$ in $\mathcal R$. The path $p''$,
however, is not necessarily m-connecting in $\mathcal R_{\underline
X}$, since it may happen that there is a collider $Q$ on the path such
that $Q \in \operatorname{an}(\mathbf{Z},\mathcal R)$ but $Q \notin
\operatorname{an}(\mathbf{Z},\mathcal R_{\underline X})$. But this can only
occur if $Q \in \operatorname{an}(X,\mathcal R_{\underline X})$. Hence, $p''$
satisfies the following properties: no noncollider on $p''$ is in
$\mathbf{Z}$ and every collider on $p''$ is in $\operatorname{an}(\mathbf{Z}
\cup\{X\}, \mathcal R_{\underline X})$.
   It then follows from Lemma~\ref{lemmaRichardsoncorollary} that there is an m-connecting path between $X$ and $Y$ given $\mathbf{Z}$ in $\mathcal R_{\underline X}$.

We now prove the ``if'' statement. Suppose that there is an m-connecting
path $p''$ between $X$ and $Y$ given $\mathbf{Z}$ in $\mathcal
R_{\underline X}$. Let $p'$ be the corresponding path in $\mathcal R$,
consisting of the same sequence of vertices. Then $p'$ is also
m-connecting given $\mathbf{Z}$ in $\mathcal R$. Moreover, $p$ does not
start with a visible edge out of $X$ in $\mathcal G$, because $p''$
exists in $\mathcal R_{\underline X}$. By Lemma~2$'$ in the proof of
Lemma~5.1.7 of \citet{Zhang06-dissertation}, it then follows that there
exists an m-connecting definite status back-door path between $X$ and
$Y$ given $\mathbf{Z}$ in $\mathcal G$.
\end{pf}

The next lemma is used several times to derive a contradiction.

\begin{lemma}\label{lemmapossiblydirectedpathandedgeinto}
Let $U$ and $V$ be two distinct vertices in $\mathcal G$, where
$\mathcal G$ denotes a DAG, CPDAG, MAG or PAG. Then $\mathcal G$ cannot
have both a possibly directed path from $U$ to $V$ and an edge of the
form $V \stararrow U$.
\end{lemma}

\begin{pf}
This lemma is trivial for DAGs and MAGs, since they cannot contain
(almost) directed cycles. So we only show the result for CPDAGs and
PAGs. Let $\mathcal G$ denote the CPDAG or PAG, and suppose that
$\mathcal G$ contains an edge of the form $V \stararrow U$ as well as a
possibly directed path from $U$ to $V$ in $\mathcal G$. Then there is
also a possibly directed definite status path $p = \langle
U=U_1,\dots,U_k=V\rangle$ from $U$ to $V$ in~$\mathcal G$, by Lemma~\ref{lemmapossibledescendantsdefinitestatuspath}. The path $p$ has the
following properties: if $U_{i-1} \stararrow U_i$ for some $i\in
\{2,\dots,k\}$, then $U_{j-1} \to U_j$ for all $j\in \{i+1,\dots,k\}$,
and the length of $p$ must be at least two, because of the edge $V
\stararrow U$.

If $p$ is fully directed, there is an (almost) directed cycle in any
DAG or MAG in the Markov equivalence class described by $\mathcal G$,
which violates the ancestral property.

Otherwise, if $p$ contains a directed sub-path, let $p(U_d,V)$ be the
longest directed sub-path. Then the sub-path $p(U,U_d)$ must be of the
form $U \circcirc \cdots\break  \circcirc U_d$ or $U \circcirc \cdots \circcirc
\circarrow U_d$. In either case, the edge $V \stararrow U$ and repeated
applications of Lemma~\ref{lemmapropertyCP1forCPDAGsandPAGs} imply the
edge $V \stararrow U_d$. This gives an (almost) directed cycle together
with the directed path $p(U_d,V)$ in any DAG or MAG in the Markov
equivalence class described by $\mathcal G$. This again contradicts the
ancestral property.

Otherwise, $p$ does not contain a directed sub-path. Let $T$ be the
vertex preceding $V$ on the path. Then the path has one of the
following two forms: $U \circcirc \cdots \circcirc T \circcirc V$ or $U
\circcirc \cdots \circcirc T \circarrow V$. The edge $V \stararrow U$
and repeated applications of Lemma~\ref{lemmapropertyCP1forCPDAGsandPAGs} yield the edge $V \stararrow T$,
which contradicts $T\circcirc V$ or $T \circarrow V$.
\end{pf}

Theorem~\ref{thbackdoorsetforgeneralgraph} requires a DAG or MAG in
$\mathcal R^*$; see Definition~\ref{defR}. The following lemma
establishes such a DAG or MAG exists, since $\mathcal R^*$ is always
nonempty. This result is closely related to constructions in
\citet{AliEtAl05}, Theorem~2 of \citet{Zhang08-orientation-rules} and
Lemma~27 of \citet{Zhang08-causal-reasoning-ancestral-graphs}.

\begin{lemma}\label{lemmaMAGwithsameindegreeasPAG}
Let $\mathcal G$ be a PAG (CPDAG) with $k$ edges into $X$, $k\in
\{0,1,\dots\}$. Then there exists at least one MAG (DAG) $\mathcal R$
in the Markov equivalence class represented by $\mathcal G$ that has
$k$ edges into $X$.
\end{lemma}

\begin{pf}
Building on the work of \citet{Meek95}, Theorem~2 of
\citet{Zhang08-orientation-rules} gives a procedure to create a MAG
(DAG) in the Markov equivalence class represented by a PAG (CPDAG)
$\mathcal G$. One first replaces all partially directed ($\circarrow$)
edges in $\mathcal G$ by directed ($\to$) edges. Next, one considers
the circle component $\mathcal G^C$ of $\mathcal G$, that is, the
sub-graph of $\mathcal G$ consisting of nondirected ($\circcirc\!$)
edges and orients this into a directed graph without directed cycles
and unshielded colliders. The first step of this procedure only creates
tail marks, and hence cannot yield an additional edge into $X$. For the
second step, we will argue that we can construct such a graph that does
not have any edges into $X$.

First, we note that $\mathcal G^C$ is chordal; that is, any cycle of
length four or more has a chord, which is an edge joining two vertices
that are not adjacent in the cycle; see the proof of Lemma~4.1 of
\citet{Zhang08-orientation-rules}. Any chordal graph with more than one
vertex has two simplicial vertices, that is, vertices $V$ such that all
vertices adjacent to $V$ are also adjacent to each other [e.g.,
\citet{Golumbic80}]. Hence, $\mathcal G^C$ must have at least one
simplicial vertex that is different from $X$. We choose such a vertex
$V_1$ and orient any edges incident to $V_1$ into $V_1$. Since $V_1$ is
simplicial, this does not create unshielded colliders. We then remove
$V_1$ and these edges from the graph. The resulting graph is again
chordal [e.g., \citet{Golumbic80}] and therefore again has at least one
simplicial vertex that is different from $X$. Choose such a vertex
$V_2$, and orient any edges incident to $V_2$ into $V_2$. We continue
this procedure until all edges are oriented. The resulting ordering is
called a perfect elimination scheme for $\mathcal G^C$. By
construction, this procedure yields an acyclic directed graph without
unshielded colliders. Moreover, since $X$ is chosen as the last vertex
in the perfect elimination scheme,
we do not orient any edges into $X$. 
\end{pf}

\begin{lemma}\label{lemmaancestorofY}
   Let $X$ and $Y$ be two distinct vertices in $\mathcal{G}$, where $\mathcal{G}$ is a DAG, CPDAG, MAG or PAG. Let
   $\mathcal R_{\underline X}$ be any graph satisfying Definition~\ref{defR}.
   If $V \in \operatorname{D\mbox{-}SEP}(X,Y,\mathcal R_{\underline X}) \cap \operatorname{possibleDe}(X,\mathcal G)$, then
   $V \in \operatorname{an}(Y,\mathcal R_{\underline X})$.
\end{lemma}

\begin{pf}
Let $\mathcal R_{\underline X}$ satisfy Definition~\ref{defR}, and let
$V \in \operatorname{D\mbox{-}SEP}(X,Y,\mathcal R_{\underline X}) \cap
\operatorname{possibleDe}(X,\mathcal G)$. This means that there is a collider
path $p_1$ between $X$ and $V$ in $\mathcal R_{\underline X}$ such that
every vertex on the path is an ancestor of $X$ or $Y$ in $\mathcal
R_{\underline X}$. In particular, $V \in \operatorname{an}(\{X,Y\}, \mathcal
R_{\underline X})$.

We first show that $V\in \operatorname{pa}(X,\mathcal R_{\underline X})$ leads
to a contradiction. Thus, suppose there is an edge $X \leftarrow V$ in
$\mathcal R_{\underline X}$. By construction of $\mathcal R_{\underline
X}$, $\mathcal G$ then contains an edge of the form $X \arrowcirc V$ or
$X \leftarrow V$, but this forms a contradiction together with $V \in
\operatorname{possibleDe}(X,\mathcal G)$, by Lemma~\ref{lemmapossiblydirectedpathandedgeinto}.

We now\vspace*{1pt} show that $V\in \operatorname{an}(X, \mathcal R_{\underline X})
\setminus \operatorname{pa}(X, \mathcal R_{\underline X})$ leads to a
contradiction. Thus suppose there is a directed path from $V$ to $X$ in
$\mathcal R_{\underline X}$ of the form $\langle V, \dots, W,
X\rangle$, where $V \neq W$ and $W\neq X$. By construction of $\mathcal
R_{\underline X}$, the edge $W \to X$ must also be into $X$ in
$\mathcal G$, so that $\mathcal G$ contains $W \circarrow X$ or $W \to
X$. Since $V \in \operatorname{possibleDe}(X,\mathcal G)$, there is a possibly
directed path $p_{xv}$ from $X$ to $V$ in $\mathcal G$. Since $\mathcal
R_{\underline X}$ contains a directed path from $V$ to $W$, $\mathcal
G$ must also contain a possibly directed path $p_{vw}$ from~$V$ to $W$.
This implies that $p_{xv} \oplus p_{vw}$ is a possibly directed path
from $X$ to $W$ in $\mathcal G$, so that $W \in \operatorname{possibleDe}(X,
\mathcal G)$. But this forms a contradiction with $W \circarrow X$ or
$W \to X$ in $\mathcal G$, by Lemma~\ref{lemmapossiblydirectedpathandedgeinto}.

Hence, we must have $V\in \operatorname{an}(Y,\mathcal R_{\underline X})$.
\end{pf}

We can now prove the main result in Section~\ref{secfindingbackdoorset}.
\begin{pf*}{Proof of Theorem~\ref{thbackdoorsetforgeneralgraph}}
Let $\mathcal R_{\underline X}$ satisfy Definition~\ref{defR}. We first
show that $Y \in \operatorname{adj}(X,\mathcal R_{\underline X})$ or
$\operatorname{D\mbox{-}SEP}(X,Y,\mathcal R_{\underline X}) \cap
\operatorname{possibleDe}(X,\mathcal G) \neq \varnothing$ implies that there
does not exist a generalized back-door set relative to $(X,Y)$ and
$\mathcal G$, since no set $\mathbf{W}$ can satisfy conditions (B-i)
and (B-ii) in Definition~\ref{defbackdoorgraphical}.
   Thus suppose that $Y \in \operatorname{adj}(X,\mathcal R_{\underline X})$. Then there is a definite status back-door path of length one in $\mathcal G$
   that cannot be blocked. Hence condition (B-ii) cannot be satisfied by any set $\mathbf{W}$.
    Next, suppose that there exists some vertex $V \in \operatorname{D\mbox{-}SEP}(X,Y,\mathcal R_{\underline X}) \cap \operatorname{possibleDe}(X,\mathcal G) \neq \varnothing$.
    Then there is a collider path $p_1$ between $X$ and $V$ in $\mathcal R_{\underline X}$ such that every vertex on the path is in $\operatorname{an}(\{X,Y\},
    \mathcal R_{\underline X})$. Moreover, by Lemma~\ref{lemmaancestorofY}, there is a directed path $p_2$ from $V$ to $Y$ in $\mathcal R_{\underline X}$.
    Now consider  $p = p_1 \oplus p_2$. All nonendpoint vertices on $p$ that are not on $p_2$ are colliders on $p$ and in $\operatorname{an}(\{X,Y\}, \mathcal R_{\underline X})$.
    The remaining nonendpoint vertices on $p$ are noncolliders and in $\operatorname{possibleDe}(X,\mathcal G)$ [since $V\in \operatorname{possibleDe}(X,\mathcal G)$],
    so that including them in $\mathbf{W}$ violates condition (B-i). It then follows by Lemma~\ref{lemmaRichardsoncorollary} that for any subset $\mathbf{W}$ satisfying
    condition (B-i), there exists an m-connecting path between $X$ and $Y$ given $\mathbf{W}$ in $\mathcal R_{\underline X}$. By Lemma~\ref{lemmam-sepPAG2MAG},
    this means that we cannot block all definite status back-door paths from $X$ to $Y$ in $\mathcal G$ without violating condition (B-i).

   We now prove the other direction. Thus\vspace*{1pt} suppose that $Y \notin
   \operatorname{adj}(X,\mathcal R_{\underline X})$ and $\operatorname{D\mbox{-}SEP}(X,Y,\mathcal
   R_{\underline X}) \cap \operatorname{possibleDe}(X, \mathcal G) = \varnothing$. Then we need to show that $\operatorname{D\mbox{-}SEP}(X, Y,\mathcal R_{\underline X})$
   satisfies conditions (B-i) and (B-ii) of Definition~\ref{defbackdoorgraphical}. Condition (B-i) is satisfied trivially, since $\operatorname{D\mbox{-}SEP}(X,Y,\mathcal R_{\underline X})
   \cap \operatorname{possibleDe}(X,\mathcal G) = \varnothing$. To prove that condition (B-ii) is satisfied as well, we first show $Y \notin \operatorname{D\mbox{-}SEP}(X,Y,\mathcal R_{\underline X})$,
   by contradiction. Thus, suppose $Y \in \operatorname{D\mbox{-}SEP}(X,Y,\mathcal R_{\underline X}) \subseteq \operatorname{D\mbox{-}SEP}(X,Y,\mathcal R)$. By Lemma~\ref{lemmaD-SEP}, this implies
   $Y \in \operatorname{adj}(X,\mathcal R)$. Since $Y\notin \operatorname{adj}(X,\mathcal R_{\underline X})$, this implies that $X\to Y$ in $\mathcal G$ with a visible edge.
   But this means that $Y\in \operatorname{possibleDe}(X,\mathcal G)$, so that $\operatorname{D\mbox{-}SEP}(X,Y,\mathcal R_{\underline X}) \cap \operatorname{possibleDe}(X,\mathcal G) \neq \varnothing$.
   This is a contradiction, which implies $Y\notin \operatorname{D\mbox{-}SEP}(X,Y,\mathcal R_{\underline X})$. Hence  $\operatorname{D\mbox{-}SEP}(X,Y,\mathcal R_{\underline X})$ m-separates $X$
   and $Y$ in $\mathcal R_{\underline X}$ by Lemma~\ref{lemmaD-SEP} (we use here that $\mathcal R_{\underline X}$ is ancestral).
   By Lemma~\ref{lemmam-sepPAG2MAG}, this implies that $\operatorname{D\mbox{-}SEP}(X,Y,\mathcal
   R_{\underline X})$ blocks all definite status back-door paths from $X$ to $Y$
   in $\mathcal G$, so that condition (B-ii) is satisfied.
\end{pf*}

\begin{pf*}{Proof of  Corollary \protect\ref{corbackdoorsetDAG}}
%
Although this result for DAGs is well known, we show how one can derive
this from Theorem~\ref{thbackdoorsetforgeneralgraph}. Note that
$\mathcal D_{\underline X}$ is the graph obtained by removing all
directed edges out of $X$ from $\mathcal D$. Moreover,
$\operatorname{D\mbox{-}SEP}(X,Y,\mathcal D_{\underline X}) = \operatorname{pa}(X,\mathcal D)$
and $\operatorname{possibleDe}(X, \mathcal D) = \operatorname{de}(X, \mathcal D)$.
   Now the condition $Y \notin \operatorname{adj}(X,\mathcal D_{\underline X})$ is equivalent to $Y\notin \operatorname{pa}(X,\mathcal D)$.
   The other condition $\operatorname{D\mbox{-}SEP}(X,Y,\mathcal D_{\underline X}) \cap
    \operatorname{possibleDe}(X,\mathcal D) = \varnothing$ reduces to $\operatorname{pa}(X,\mathcal D) \cap \operatorname{de}(X,\mathcal D)=\varnothing$,
    and this is fulfilled automatically by the acyclicity of $\mathcal D$. Hence Theorem~\ref{thbackdoorsetforgeneralgraph} reduces to the given statement.
\end{pf*}

\begin{pf*}{Proof of Corollary \protect\ref{corbackdoorsetCPDAG}}
     Let $\mathcal D$ be a DAG in the Markov equivalence class represented by $\mathcal C$, constructed without orienting additional edges
     into $X$. Let $\mathcal D_{\underline X}$ be obtained from $\mathcal D$ by removing all directed edges out of $X$ that were
     directed out of $X$ in $\mathcal C$. Let $\mathcal C_{\underline X}$ be obtained from $\mathcal C$ by removing all directed edges out of $X$.

We first show that $Y \in \operatorname{pa}(X, \mathcal C)$ or  $Y \in
\operatorname{possibleDe}(X,\mathcal C_{\underline X})$ imply $Y\in
\operatorname{adj}(X, \mathcal D_{\underline X})$ or $\operatorname{D\mbox{-}SEP}(X,Y,\mathcal
D_{\underline X}) \cap \operatorname{possibleDe}(X,\mathcal C) \neq
\varnothing$. Thus suppose $Y \in \operatorname{pa}(X, \mathcal C)$. Then $Y\in
\operatorname{adj}(X,\mathcal D_{\underline X})$.
    Next, suppose $Y \in \operatorname{possibleDe}(X,\mathcal C_{\underline X})$. It can be easily shown that $\mathcal C_{\underline X}$
    satisfies the basic property of Lemma~\ref{lemmapropertyCP1forCPDAGsandPAGs}, that $A \to B \circcirc C$ implies $A\to C$ (since all edges that are
removed are directed edges out of $X$). Hence, Lemma~\ref{lemmapossibledescendantsdefinitestatuspath} applies to $\mathcal
C_{\underline X}$, and it follows that there is a possibly directed
definite status path from $X$ to $Y$ in $\mathcal C_{\underline X}$.
All nonendpoint vertices on this path must be definite noncolliders.
By construction of $\mathcal C_{\underline X}$, the first edge on this
path must be nondirected in $\mathcal C_{\underline X}$, and by
construction of $\mathcal D_{\underline X}$, this edge must be oriented
out of $X$ in $\mathcal D_{\underline X}$. This implies that the entire
path must be directed from $X$ to $Y$ in $\mathcal D_{\underline X}$,
since all nonendpoint vertices are noncolliders. Let $V$ be the
vertex adjacent to $X$ on the path. Then $V\in
\operatorname{D\mbox{-}SEP}(X,Y,\mathcal D_{\underline X})$. Moreover, $V\in
\operatorname{possibleDe}(X, \mathcal C)$. Hence $\operatorname{D\mbox{-}SEP}(X,Y,\mathcal
D_{\underline X}) \cap \operatorname{possibleDe}(X,\mathcal C) \neq
\varnothing$.

We now show that $\operatorname{D\mbox{-}SEP}(X,Y,\mathcal D_{\underline X})
 \cap
\operatorname{possibleDe}(X,\mathcal C) \neq \varnothing$ or $Y\in\break 
\operatorname{adj}(X, \mathcal D_{\underline X})$ imply $Y \in \operatorname{pa}(X,
\mathcal C)$ or  $Y \in \operatorname{possibleDe}(X,\mathcal C_{\underline
X})$. Thus\vspace*{1pt} suppose that $Y \in \operatorname{pa}(X, \mathcal C)$ imply $Y \in
\operatorname{pa}(X, \mathcal C)$ or  $Y \in \operatorname{possibleDe}(X,\mathcal
C_{\underline X})$. Thus suppose $Y \in \operatorname{adj}(X,\mathcal
D_{\underline X})$. Then either $X \leftarrow Y$ or $X \circcirc Y$ in
$\mathcal C$. This implies that $Y\in \operatorname{pa}(X,\mathcal C)$ or $Y\in
\operatorname{possibleDe}(X,\mathcal C_{\underline X})$. Next, suppose that
there exists a vertex $V \in \operatorname{D\mbox{-}SEP}(X,Y, \mathcal D_{\underline
X}) \cap \operatorname{possibleDe}(X,\mathcal C)$. Note that $V \in
\operatorname{D\mbox{-}SEP}(X,\break Y, \mathcal D_{\underline X})$ implies: (i) $V\in
\operatorname{pa}(X,\mathcal D_{\underline X})$ or (ii) $V\in
\mbox{ch}(X,\mathcal D_{\underline X}) \cap \operatorname{an}(Y,\mathcal
D_{\underline X})$ or (iii) $V\in \operatorname{pa}( \mbox{ch}(X,\mathcal
D_{\underline X}) \cap \operatorname{an}(Y,\mathcal D_{\underline X}))$. By
construction of $\mathcal D_{\underline X}$, case (i) implies $V\in
\operatorname{pa}(X,\mathcal C)$. But this is in contradiction with $V\in
\operatorname{possibleDe}(X,\mathcal C)$, by Lemma~\ref{lemmapossiblydirectedpathandedgeinto}. In case (ii), we have $X
\to V$ and a directed path from $V$ to $Y$ in $\mathcal D_{\underline
X}$, so that $Y \in \operatorname{de}(X,\mathcal D_{\underline X})$. Similarly,
we can obtain $Y \in \operatorname{de}(X,\mathcal D_{\underline X})$ in case
(iii). This implies $Y \in \operatorname{possibleDe}(X,\mathcal C_{\underline
X})$ in cases (ii) and (iii).

The above shows the following: if $Y\in \operatorname{pa}(X,\mathcal C)$ or
$Y\in \operatorname{possibleDe}(X,\mathcal C_{\underline X})$, then it is
impossible to satisfy the generalized back-door criterion relative to
$(X,Y)$ and $\mathcal C$. On the other hand, if $Y \notin
\operatorname{pa}(X,\mathcal C)$ and $Y\notin \operatorname{possibleDe}(X,\mathcal
C_{\underline X})$, then $\operatorname{D\mbox{-}SEP}(X,Y,\mathcal D_{\underline X})$
satisfies the generalized back-door criterion relative to $(X,Y)$ and
$\mathcal C$. It is left to show that in the latter case, we can
replace $\operatorname{D\mbox{-}SEP}(X,Y,\mathcal D_{\underline X})$ by
$\operatorname{pa}(X,\mathcal C)$. Since $\operatorname{pa}(X,\mathcal C) \subseteq
\operatorname{D\mbox{-}SEP}(X,Y,\mathcal D_{\underline X})$, it is clear that
$\operatorname{pa}(X,\mathcal C)$ satisfies condition (B-i) of Definition~\ref{defbackdoorgraphical}. We will now show that it also satisfies
condition (B-ii).

Thus, suppose that $Y\notin \operatorname{pa}(X,\mathcal C)$ and $Y \notin
\operatorname{possibleDe}(X,\mathcal C_{\underline X})$. Consider a definite
status back-door path $p = \langle X=U_1,\dots,U_k=Y\rangle$ from $X$
to $Y$ in $\mathcal C$. Since $p$ is a back-door path, it must start
with $X \leftarrow U_2$ or $X\circcirc U_2$. Moreover, the length of
$p$ is at least two. If $X \leftarrow U_2$, then it is clear that
$\operatorname{pa}(X,\mathcal C)$ blocks $p$. If $X \circcirc U_2$, then $p$
cannot have a sub-path of the form $U_{i-1} \circcirc U_i \leftarrow
U_{i+1}$, $i\in \{2, \dots, k-1\}$, because $U_i$ is of a definite
status. Moreover, $p$ cannot be possibly directed, because then $Y \in
\operatorname{possibleDe}(X,\mathcal C_{\underline X})$. Hence, there must be
at least one collider on $p$.  Let $Q$ be the collider on $p$ that is
closest to $X$. Then the sub-path $p(X,Q)$ is a possibly directed path
from $X$ to $Q$ in $\mathcal C$. Suppose that $Q$ is an ancestor of
some vertex $W \in \operatorname{pa}(X,\mathcal C)$ in $\mathcal C$. Then there
is a possibly directed path from $X$ to $W$ in $\mathcal C$, as well as
an edge $W \to X$. But this is impossible by Lemma~\ref{lemmapossiblydirectedpathandedgeinto}. Hence, $Q$ cannot be an
ancestor of any member of $\operatorname{pa}(X,\mathcal C)$ in $\mathcal C$.
This implies that $p$ is blocked by $\operatorname{pa}(X,\mathcal C)$.
\end{pf*}

\section*{Acknowledgements}

We are very grateful to Markus Kalisch, Thomas Richardson and two anonymous referees for their
comments and suggestions that have significantly improved the paper.



\printaddresses
\end{document}